\documentclass[fleqn,usenatbib]{mnras}

\usepackage{newtxtext,newtxmath}

\usepackage[T1]{fontenc}
\usepackage{ae,aecompl}

\usepackage{graphicx}	
\usepackage{amsmath}	
\usepackage{amssymb}	
\usepackage{threeparttable}
\usepackage{lscape}
\usepackage{pdflscape}
\usepackage[separate-uncertainty=true,multi-part-units=single]{siunitx}
\usepackage[strings]{underscore}
\usepackage{titlesec}

\DeclareSIUnit\jansky{Jy}
\DeclareSIUnit\beam{PSF}
\newcommand{\mjpb}{\milli\jansky\per\beam}
\newcommand{\starname}[1]{\vspace*{4pt}\noindent {\bf #1}}

\title[RACS Stokes V Stars]{A circular polarisation survey for radio stars with the Australian SKA Pathfinder}

\author[Joshua Pritchard et al.]
{
Joshua Pritchard,$^{1,2,3}$\thanks{Email: joshua.pritchard@sydney.edu.au}
Tara Murphy,$^{1,3}$\thanks{Email: tara.murphy@sydney.edu.au}
Andrew Zic,$^{1,2}$
Christene Lynch,$^{4,5}$
\newauthor{}
George Heald,$^{6}$
David L.~Kaplan,$^{7}$
Craig Anderson,$^{8,2}$
Julie Banfield,$^{2}$
Catherine Hale,$^{6}$
\newauthor{}
Aidan Hotan,$^{6}$
Emil Lenc,$^{2}$
James K. Leung,$^{1,2,3}$
David McConnell,$^{2}$
Vanessa A. Moss,$^{2,1}$
\newauthor{}
Wasim Raja,$^{2}$
Adam J.~Stewart,$^{1}$
and Matthew Whiting$^{2}$
\\
$^{1}$Sydney Institute for Astronomy, School of Physics, University of Sydney, NSW 2006, Australia\\
$^{2}$CSIRO Astronomy and Space Science, PO Box 76, Epping, NSW 1710, Australia\\
$^{3}$ARC Centre of Excellence for Gravitational Wave Discovery (OzGrav), Hawthorn, Victoria, Australia \\
$^{4}$International Centre for Radio Astronomy Research (ICRAR), Curtin University, Bentley, WA, Australia \\
$^{5}$ARC Centre of Excellence for All Sky Astrophysics in 3 Dimensions (ASTRO3D), Bentley, WA, Australia \\
$^{6}$CSIRO Astronomy and Space Science, PO Box 1130, Bentley, WA 6102, Australia \\
$^{7}$Department of Physics, University of Wisconsin--Milwaukee, Milwaukee, Wisconsin 53201, USA.\\
$^{8}$National Radio Astronomy Observatory, 1003 Lopezville Rd, Socorro, NM 87801, USA
}

\date{Accepted XXX. Received YYY; in original form ZZZ}

\pubyear{2021}

\begin{document}
\label{firstpage}
\pagerange{\pageref{firstpage}--\pageref{lastpage}}
\maketitle

\begin{abstract}
  We present results from a circular polarisation survey for radio stars in the Rapid ASKAP
  Continuum Survey (RACS). RACS is a survey of the entire sky south of $\delta=+41^\circ$
  being conducted with the Australian Square Kilometre Array Pathfinder telescope
  (ASKAP) over a \SI{288}{\mega\hertz} wide band centred on \SI{887.5}{\mega\hertz}. The data
  we analyse includes Stokes I and V polarisation products to an RMS sensitivity of
  \SI{250}{\micro\jansky\per\beam}. We searched RACS for sources with fractional circular
  polarisation above 6 per cent, and after excluding imaging artefacts, polarisation leakage,
  and known pulsars we identified radio emission coincident with 33 known stars. These range
  from M-dwarfs through to magnetic, chemically peculiar A- and B-type stars. Some of these are
  well known radio stars such as YZ CMi and CU Vir, but 23 have no previous radio
  detections. We report the flux density and derived brightness temperature of these detections
  and discuss the nature of the radio emission. We also discuss the implications of our results
  for the population statistics of radio stars in the context of future ASKAP and Square
  Kilometre Array surveys.
\end{abstract}

\begin{keywords}
radio continuum: stars -- stars: low mass -- stars: chemically peculiar
\end{keywords}

\section{INTRODUCTION}

While most stars generally have weak radio emission and are undetectable over Galactic distance
scales, magnetically active stars are known to exhibit radio bursts with intensities orders of
magnitude greater than those produced by the Sun. Radio bursts have been detected from a wide
range of stellar types, including chromospherically active M-dwarfs, ultracool dwarfs, close
and interacting binaries, and magnetic chemically peculiar (MCP) stars (see \citealt{Gudel2002}
for a review). Stellar radio bursts are often highly circularly-polarised with brightness
temperatures in excess of \SI{e12}{\kelvin}, requiring the operation of a non-thermal, coherent
emission process. Coherent radio bursts are generally attributed to one of two
processes:\,plasma emission which operates at the fundamental and second harmonic of the local
plasma frequency $\omega_p=\sqrt{4\pi{n_e}e^2/m_e}$, or electron cyclotron maser (ECM) emission
at the local relativistic cyclotron frequency $\omega_c=eB/\gamma{m_e}c$ (see
\citealt{Dulk1985} for a detailed review of these processes). Here $n_e$, $e$, and $m_e$ are
the electron density, charge, and mass respectively, $B$ is the magnetic field strength,
$\gamma$ is the Lorentz factor, and $c$ is the speed of light.

If the emission mechanism can be determined, these processes are an excellent measure of either
the electron density or magnetic field of the stellar magnetosphere, and provide a means to
probe extreme events such as particle acceleration driven by magnetic reconnection
\citep{Crosley2016} and auroral current systems \citep{Leto2017}. Periodic auroral bursts are
often highly beamed and modulated by stellar rotation, providing an independent
constraint on the stellar rotation period \citep{Zic2019}. Constraints on the energetics and
rates of stellar radio bursts also inform models of magnetospheric topology \citep{Andre1991},
cool star magnetic dynamos \citep{Kao2016}, the origin of strong magnetic fields in hot stars
\citep{Schneider2016}, and the habitability of exoplanets \citep{Crosley2018a}.

Despite the wide variety of stars demonstrating non-thermal radio bursts, studies have
typically been limited to targeted observations of a small number of candidates, whose
selection is motivated by activity indicators such as prior optical or X-ray flaring
\citep{White1989b, Gudel1992}, presence of chromospheric emission lines \citep{Slee1987}, or
previously known radio activity \citep{Villadsen2019}. Surveys have also been conducted
specifically targeting known classes of radio star such as RS Canum Venaticorum (RS CVn) and
Algol binaries \citep{Collier1982, Morris1988, Umana1998}, hot OB-type stars
\citep{Bieging1989}, young stellar objects (YSOs) \citep{Andre1987, Osten2009}, and ultracool
dwarfs \citep{Antonova2013}. The selection biases inherent in these targeted searches impacts
the inference of population statistics, and does not allow for the discovery of new classes of
radio stars.

A few volume-limited surveys for radio stars have been conducted, though their success has been
limited by the high surface density of background radio sources, resulting in a large number of
false-positive matches. \citet{Helfand1999} searched \SI{5000}{\deg^2} of high Galactic
latitude sky to a sensitivity of \SI{0.7}{\mjpb} in the VLA Faint Images of the Radio Sky at
Twenty-cm \citep[FIRST;][]{Becker1995} survey identifying 26 radio stars. \citet{Kimball2009}
further explored FIRST in a comparison with the Sloan Digital Sky Survey
\citep[SDSS;][]{Adelman-McCarthy2008} selecting 112 point radio sources coincident with
spectrally confirmed SDSS stars, though a similar number of matches are estimated due to chance
alignment with background galaxies. \citet{Umana2015} surveyed a \SI{4}{\deg^2} field located
within the Galactic plane to a sensitivity of \SI{30}{\micro\jansky\per\beam} in a search for
stellar radio emission, and identified 10 hot stars producing thermal radio emission out of 614
detected sources. \citet{Vedantham2020a} crossmatched radio sources in the LOFAR Two-Metre Sky
Survey \citep[LOTSS;][]{Shimwell2017} with nearby stars in Gaia Data Release 2
\citep{Andrae2018}, discovering an M-dwarf producing coherent, circularly polarised auroral
emission at metre wavelengths.

Circular polarisation surveys are a promising method for wide-field detection of stellar radio
bursts, as the synchrotron emission from Active Galactic Nuclei (AGN) that accounts for the
majority of unresolved radio sources is typically less than $1-2$ per cent circularly-polarised
\citep{Macquart2002}, and thus the number of false-positive matches between stars and unrelated
background AGN is significantly reduced. \citet{Lenc2018} performed an all-sky circular
polarisation survey at \SI{200}{\mega\hertz} with the Murchison Widefield Array
\citep[MWA;][]{Bowman2013} detecting 33 previously known pulsars and two magnetically-active
stars. There have been no other all-sky circular polarisation radio surveys, and none have been
conducted at centimetre wavelengths where targeted campaigns of magnetically-active stars have
previously been focused \citep[e.g.,][]{Antonova2013, Villadsen2019}.

The Australian Square Kilometre Array Pathfinder \citep[ASKAP;][Hotan et al. {\it in
  press}]{Johnston2008} is a radio telescope array of 36$\times$12-metre antennas located in
the Murchison Radio-astronomy Observatory in Western Australia. As part of the Rapid ASKAP
Continuum Survey \citep[RACS;][]{McConnell2020} we have performed the first all-sky circular
polarisation survey for radio stars at centimetre wavelengths. In \autoref{sec:data} we
describe the RACS observations and data processing. In \autoref{sec:search} we describe our
search procedure for stars with significant circular polarisation. We present the detected
radio stars in \autoref{sec:results} and discuss the implications of our survey on stellar
radio burst statistics and future ASKAP surveys in \autoref{sec:discussion}.

\section{OBSERVATIONS AND DATA REDUCTION}\label{sec:data}

RACS is the first all-sky survey performed with the full 36 antenna ASKAP telescope, and covers
the entire sky south of $\delta=+41^\circ$. Each antenna in the array is equipped with a Phased
Array Feed \citep[PAF;][]{Hotan2014, McConnell2016} which allows 36 dual linear polarisation
beams to be formed on the sky. All four cross-correlations were recorded, allowing full
  Stokes I, Q, U and V images to be reconstructed.  The antenna roll-axis was adjusted
  throughout the observations to maintain orientation of the linear feeds with respect to the
  celestial coordinate frame, so the beam footprint remained fixed on the sky and no correction
  for parallactic angle was required. RACS observations used a 6$\times$6-beam square
footprint giving a \SI{32}{\deg^2} field of view, and were acquired in 15 minute integrations
at a central frequency of \SI{887.5}{\mega\hertz} with 288 channels each \SI{1}{\mega\hertz}
wide. RACS observations were conducted from April through May 2019, with fields originally
located near the Sun observed in August 2019. 

\subsection{Data Reduction}\label{sec:data-reduction}

ASKAP data is calibrated and imaged with the {\sc ASKAPsoft} package \citep{Cornwell2011,
  Guzman2019} using the {\em Galaxy} computer cluster that is maintained at the Pawsey
Supercomputing Centre.  The specific methods used for RACS data reduction are described in
detail by \citet{McConnell2020}, but we summarise them here.

Data were prepared for imaging by flagging bad samples and applying a calibration derived from
the calibration source PKS\,B1934$-$638, which included factors for adjustment of
interferometric phase and for setting the absolute flux density scale. Images were formed for
each beam as the first two terms (0, 1) of a Taylor series in frequency using multi-frequency
synthesis. {\sc ASKAPsoft} applies weights to the visibility data using {\em preconditioning}
\citep{Rau2010}; RACS data were weighted to achieve the equivalent of a~\cite{Briggs1995}
`robust weight' of $r = 0.0$, where $r$ takes values between $-2$ and $2$ corresponding to
uniform and natural weighting respectively. All beam images were deconvolved using the {\sc
  BasisfunctionMFS} algorithm before being combined into a single image by linear
mosaicing. Beam images were made with 6144 $\times$ 6144 pixels each of size $2\farcs5$, and
the final mosaics have approximately\footnote{The mosaic boundary is determined by the values
  of primary beam weights and is slightly irregular.} 11800 $\times$ 11800 pixels of the same
size.

\subsection{Data Quality}\label{sec:data-quality}

A complete description of the published RACS images and source catalogue, including final 
quality control metrics, are presented by \citet{McConnell2020} and Hale et al. 
({\it in prep.})  respectively. We describe the main quality metrics relevant to an early 
processing of the data used in this paper.  

The mosaic images have a point-spread-function (PSF) that varies over the field of view and
between images due to the variation in sampling of the ($u$,$v$)-plane by individual beams. The
central lobe of the PSF has a typical major axis full width half maximum (FWHM) $B_{maj}$ of
$18\farcs0$ and minor axis FWHM $B_{min}$ of $11\farcs8$ with variations of $4\farcs3$ and
$0\farcs9$ respectively. We assessed the flux density uncertainty introduced by PSF variation
by direct comparison to the RACS source catalogue presented in Hale et al. ({\it in
  prep.}). This catalogue was produced from images that are convolved to a common resolution of
\SI{25}{arcsec} with a uniform PSF, and has a median flux density ratio of
$S_{\text{RACS}}/S_{\text{SUMSS}}\approx 1 \pm 0.20$ derived from comparison to the Sydney
University Molonglo Sky Survey \citep[SUMSS;][]{Mauch2003}.  We selected a sample of bright,
unresolved RACS catalogue sources which are isolated from neighbouring components by at least
$150\arcsec$ such that they are free from contamination due to close neighbours, and
crossmatched this sample with their counterparts in our images. We find a median ratio between
our fluxes and those from the RACS source catalogue of $1.03\pm0.09$.  The uncertainty in this
ratio contains contributions from random errors attributable to signal to noise ratio (SNR) and
position-dependent effects described by \citet{McConnell2020}. We scaled our fluxes by this 3
per cent factor and incorporated the 9 per cent uncertainty in quadrature with the RACS
catalogue flux scale uncertainty, arriving at a cumulative flux density uncertainty of
$\Delta S/S=0.22$.
  
Our positional accuracy was assessed by comparison to the second realisation of the
International Celestial Reference Frame \citep[ICRF2;][]{Fey2015} catalogue. Although ICRF2
sources are sparsely distributed across the sky, 2059 are a good positional match to the
bright, unresolved, and isolated sample described above. The distribution of positional offsets
between RACS and ICRF2 for these sources has a systematic shift of $-0\farcs6$ in right
ascension and $+0\farcs1$ in declination, and standard deviation of $0\farcs5$ in both
coordinates. This positional uncertainty is less than the pixel scale of $2\farcs5$.

The images that we analyse in this paper were made before the development of the on-axis
polarisation leakage correction described by \citet{McConnell2020}. We assessed the magnitude
of leakage between Stokes I and V by assuming all field sources with peak total intensity flux
density ${S_{888}>\SI{300}{\mjpb}}$ are unpolarised, such that any observed Stokes V flux is
solely due to leakage. A small number of pulsars are brighter than this threshold and
intrinsically polarised, though the source density of bright AGN is much greater allowing for a
good estimate of the magnitude of leakage. Among 11974 field sources in negative Stokes V and
1217 in positive\footnote{Throughout this paper we adopt the IAU sign convention for
  which positive Stokes V corresponds to right handed circular polarisation and negative Stokes
  V to left handed, as seen from the perspective of the observer.}, we find a median
polarisation leakage of $0.65$ and $0.54$ per cent respectively, with hot spots of up to
${\sim}2$ per cent near the mosaic corners.\footnote{The asymmetry between the number 
of negative and positive components is due to a bias in the Stokes V leakage pattern, and is 
currently being investigated. The real, polarised sources detected in this paper are evenly 
distributed between positive and negative.}

\begin{table}
  \centering
  \caption{Summary of early processing RACS data properties.}\label{tab:racs-properties}
    \begin{tabular}[h!]{lr}
      \hline
      Property                         & Value \\
      \hline
      Sky Coverage                     & $-90^\circ \leq \delta < +41^\circ$ \\
      Central Frequency                & \SI{887.5}{\mega\hertz} \\
      Bandwidth                        & \SI{288}{\mega\hertz} \\
      Integration Time                 & \SI{15}{\minute} \\
      Typical RMS Noise                & \SI{250}{\micro\jansky\per\beam} \\
      Cumulative Flux Density Accuracy & $\Delta S/S = 0.22$ \\
      Polarisation Leakage             & $0.65\%$ (V negative) \\
                                       & $0.54\%$ (V positive) \\
      Astrometric Accuracy             & $\Delta\alpha\cos{\delta} = -0\farcs6 \pm 0\farcs5$ \\
                                       & $\Delta\delta = +0\farcs1 \pm 0\farcs5$ \\
      PSF Central Lobe (FWHM)          & $B_{min} = 11\farcs8 \pm 0\farcs9$ \\
                                       & $B_{maj} = 18\farcs0 \pm 4\farcs3$ \\
      \hline
    \end{tabular}
\end{table}

The properties of the early processing data analysed in this paper are summarised in
Table~\ref{tab:racs-properties}. Further refinements have been applied to the published RACS
images and source catalogue, and we refer the reader to \citet{McConnell2020} and Hale et
al. ({\it in prep.}) for a complete description of the publicly available data.
  
\section{CANDIDATE SEARCH}\label{sec:search}

We used the {\sc selavy} \citep{Whiting2012b} source finder package with default settings to
extract source components from the Stokes V images, with source extraction run once to extract
$39,553$ positive components and a second time on the inverted images to extract $99,647$
negative components. We crossmatched the combined $139,200$ polarised components against the
extracted Stokes I components using a $2\arcsec$ match radius. This match radius was chosen to
avoid contamination from matches between sidelobes of bright Stokes I components and the
leakage of the bright central component into Stokes V, though it is restrictive as positional
offsets between Stokes I and V up to ${\sim} 5\arcsec$ exist for genuine circularly polarised
source components. We selected this restricted sample to investigate robust methods to reject
imaging artefacts in future searches, and to reduce the search volume to a reasonable size.

{\sc selavy} models source components with 2D Gaussians that have an SNR dependent positional
error \citep{Condon1997} of
\begin{equation}
  \sigma_{\theta} = \frac{2\theta_m}{\text{SNR}\sqrt{8\ln{2}}}
\end{equation}
where $\theta_m$ is the component major axis. We added this uncertainty in quadrature to the
astrometric accuracy discussed in \autoref{sec:data-quality} to determine the positional
uncertainty of each radio source, where the positional uncertainty corresponding to a $5\sigma$
component with major axis equal to the typical PSF major axis of $18\farcs0$ is
approximately $3\farcs2$.

\begin{figure}
\centering
\includegraphics[width=0.48\textwidth]{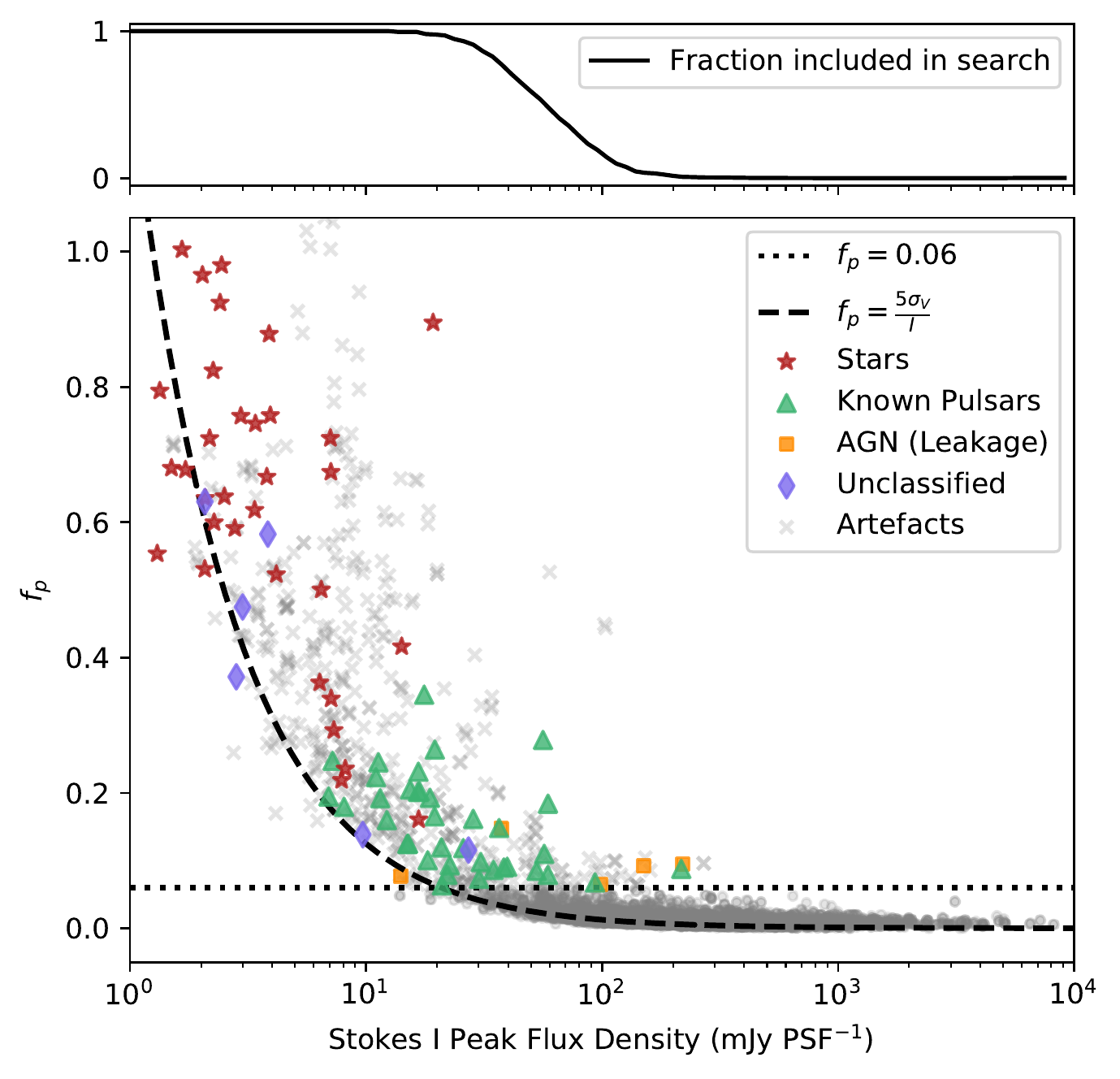}
\caption{\small Classification of visually inspected candidates. The dotted line indicates a
  fractional circular polarisation of 6 per cent below which we excluded all source components,
  labelled as grey circles. The dashed line indicates a $5\sigma_V$ detection threshold where
  $\sigma_V=$~\SI{0.25}{\mjpb} is the typical RMS noise in the Stokes V data, and represents
  the minimum fractional polarisation our search is sensitive to. Imaging artefacts are
  labelled as grey crosses, and polarisation leakage of bright AGN as yellow squares. The radio
  stars identified in our sample are labelled as red stars, and known pulsars as green
  triangles. Six circularly polarised sources of unknown classification are labelled as blue
  diamonds. The top panel shows the fraction of all components selected for visual inspection as
  a function of flux density.}\label{fig:classifications}
\end{figure}

We selected 850 components with fractional polarisation $f_p = |V|/I$ greater than $6$ per cent
for visual inspection. This corresponds to 10 times the median circular polarisation leakage,
and three times the excess leakage of ${\sim} 2$ per cent near mosaic corners. The majority of
selected components were either imaging artefacts caused by leakage of bright Stokes I sidelobes,
spurious noise near the edge of the image where sensitivity decreases, or components with
extended structure associated with radio galaxies, which we excluded by visual inspection of
the Stokes V and total intensity images. To aid further classification of our sample, we
inspected image cutouts for the SUMSS, second epoch Molonglo Galactic Plane Survey
\citep[MGPS-2;][]{Murphy2007}, NRAO VLA Sky Survey \citep[NVSS;][]{Condon1998}, TIFR GMRT Sky
Survey \citep[TGSS;][]{Intema2017}, GaLactic and Extragalactic All-sky MWA
\citep[GLEAM;][]{Wayth2015, Hurley-Walker2017}, and VLA Sky Survey \citep[VLASS;][]{Lacy2020}
radio surveys, and produced radio spectral energy distributions (SEDs) and light curves for
each candidate using the corresponding radio source catalogues, or $3\sigma$ upper limits in
the case of a non-detection.

To identify optical and infra-red counterparts to our candidates we generated image cutouts
from the \textit{Widefield Infra-red Survey Explorer} \citep[WISE;][]{Wright2010}, 2-Micron All
Sky Survey \citep[2MASS;][]{Skrutskie2006}, Panoramic Survey Telescope and Rapid Response
System 1 \citep[Pan-STARRS1;][]{Chambers2016}, and Skymapper \citep{Onken2019} surveys. As very
few stars are persistent radio sources, detection in multiple other radio surveys is suggestive
of emission from background AGN. We calculated the WISE infra-red colours \citep[see
fig. 12,][]{Wright2010} to distinguish these cases, as infra-red sources associated with
background galaxies are typically reddened compared to stars. We also looked for evidence of
the persistent X-ray and ultraviolet emission associated with some active stars in the
\textit{ROSAT} All Sky Survey \citep[RASS;][]{Boller2016} and \textit{Galaxy Evolution
  Explorer} \citep[GALEX;][]{Morrissey2007} surveys.

We excluded five components that were a positional match for previously known radio sources
attributed to bright AGN, either due to the shape of the radio SED, the WISE colours, or
explicit identification in the literature. We attribute the Stokes V flux density for these
sources to polarisation leakage as they are located near the mosaic edge at $\delta=+41^\circ$
where leakage increases, and AGN typically display fractional polarisation of at most 2 per
cent \citep{Macquart2002}. We identified 37 components as known pulsars, and six significantly
polarised sources without any multi-wavelength counterpart or catalogued identification, which
will be the subject of separate publications.

We queried the SIMBAD and NED databases for stars within a $3\farcm5$ radius of candidate
positions, which is sufficient to account for motion of the highest known proper motion stars
over a 20 year period. We applied positional corrections to all query results with available
proper motion parameters from Gaia Data Release 2, and identified 33 components that were a
match with stars to within our positional uncertainty. Our classifications are summarised in
Fig.~\ref{fig:classifications} where we show the fractional polarisation of all extracted
components as a function of $S_{888}$. Below $S_{888}=\SI{20.8}{\mjpb}$ the minimum detectable
fractional polarisation is determined by the $5\sigma$ detection limits in Stokes V rather than
contamination from polarisation leakage or the $f_p>0.06$ filter we applied.

We quantified the false-positive rate for chance-alignment between our sources and unrelated
stars by offsetting the positions of all $139,200$ Stokes V components, inclusive of artefacts,
in random directions by $5-20^\circ$ and crossmatching the new positions against the SIMBAD
database. This procedure produced no matches within $2\arcsec$, suggesting a false-positive
rate of less than $2.2\times 10^{-5}$ and an extremely low probability that any of our
detections are due to chance alignment with background stars.

\section{RESULTS}\label{sec:results}

\begin{figure}
\centering
\includegraphics[width=0.48\textwidth]{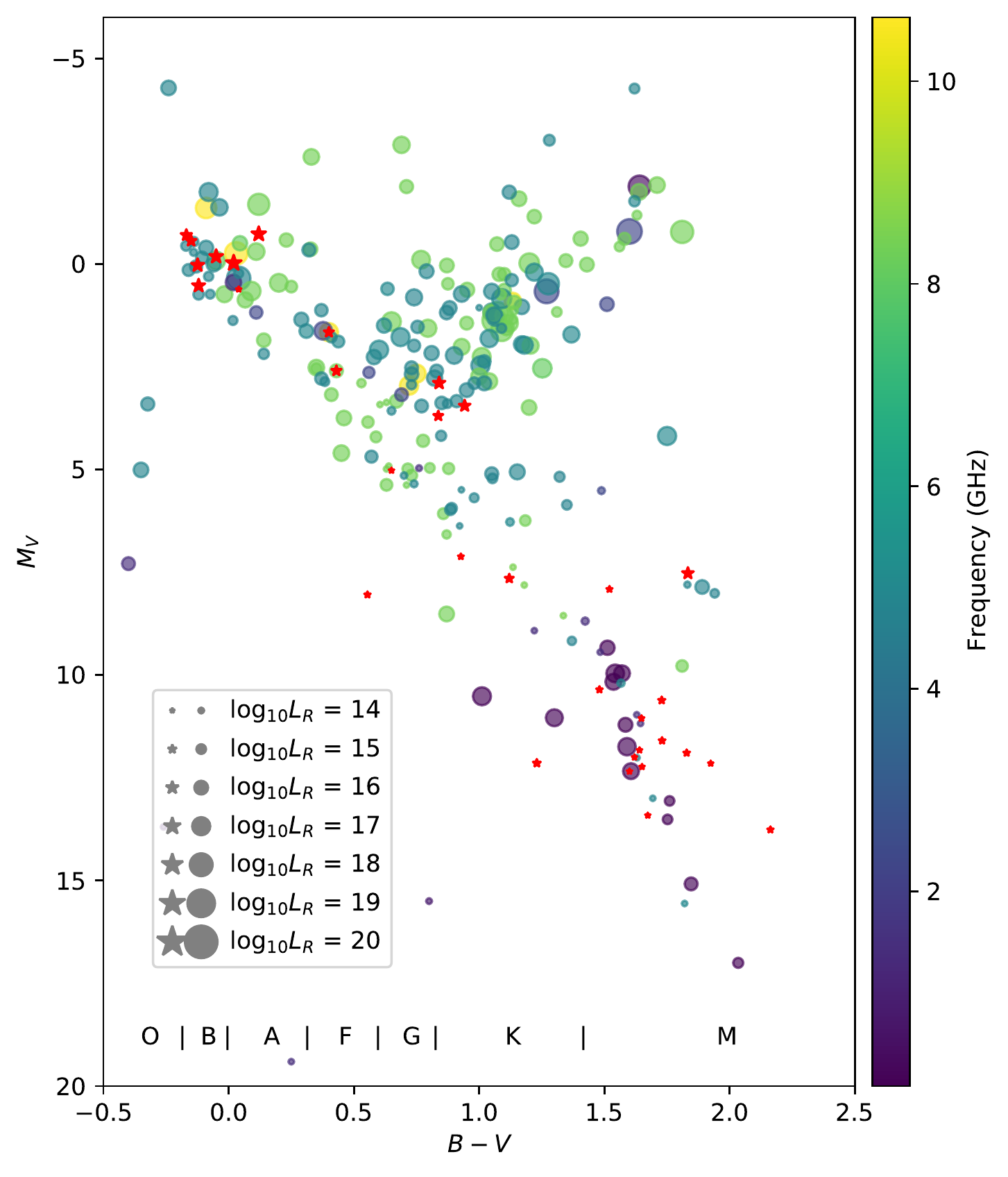}
\caption{\small Radio-selected Hertzprung-Russell diagram, showing 231 previously detected
  radio stars \citep{Wendker1995} as circles with our sample overlaid as red stars. The size of
  each marker represents the greatest radio luminosity recorded for that star, and the
  colour-map indicates the observing frequency.}\label{fig:radio-hr}
\end{figure}

Table~\ref{table:radiostars} lists the 33 radio stars detected within our sample, 23 of which
have not previously been reported as radio-loud in the literature. Fig.~\ref{fig:radio-hr}
shows a radio-selected Hertzprung-Russell diagram with 231 stars from the Wendker catalogue
\citep{Wendker1995} colour-mapped by observing frequency. Our detections are overlaid as red
stars and span the diagram, including magnetic chemically peculiar stars, young stellar
objects, RS CVn and Algol binaries, and both chromospherically active and non-active K- and
M-dwarfs. Fig.~\ref{fig:cutouts1}--\ref{fig:cutouts7} show cutout images in Stokes I and Stokes
V for each of the stars in our sample, and optical data from Pan-STARRS1 or Skymapper at
declinations above and below $-30^\circ$ respectively. We have applied astrometric corrections
to the optical data to the RACS epoch to account for proper motion, and all images are centred
on the radio position.

\begin{landscape}
  \begin{table}
    \centering \caption{Table of detected radio stars. Columns are stellar name, spectral class,
  radio coordinates, assumed upper limit to emission region length scale $R_e$, stellar parallax,
  \SI{887.5}{\mega\hertz} Stokes I peak flux density $S_{888}$, signed fractional polarisation, lower limit to
  brightness temperature, and previous radio detection references.}\label{table:radiostars}
\resizebox{\columnwidth}{!}{
\begin{threeparttable}
\begin{tabular}{llrrllccccp{3.7cm}}
\hline
 \multicolumn{1}{c}{Name}   & \multicolumn{1}{c}{Spectral Class}   & \multicolumn{1}{c}{RA (J2000)}   & \multicolumn{1}{c}{Dec (J2000)}   & \multicolumn{1}{c}{$R_e (R_\odot)$\tnote{\textdagger}}   & \multicolumn{1}{c}{$R_e$ Ref.}   & \multicolumn{1}{c}{$\pi$ (mas)}   & \multicolumn{1}{c}{$S_{888}$ (\si{\mjpb})}   & \multicolumn{1}{c}{$f_p$\tnote{*}}   & \multicolumn{1}{c}{$\log_{10}T_B$}   & \multicolumn{1}{c}{Radio Ref.}   \\
\hline
\multicolumn{10}{c}{Cool Dwarfs} \\
\hline
 G~131--26                  & M5V                                  & 00:08:53.89                      & $+$20:50:19.39                    & $~\,0.75 \pm 0.27$                                       &                                  & $~\,~\,55.3 \pm 0.8$              & $~\,~\,2.68 \pm 1.17$                        & $~\,0.594 \pm 0.211$                 & $~\,10.62 \pm 0.24$                  &                                  \\
 CS~Cet                     & K0IV(e)                              & 01:06:48.93                      & $-$22:51:23.20                    & $11.10 \pm 0.27$                                         &                                  & $~\,~\,10.5 \pm 0.1$              & $~\,~\,6.19 \pm 1.50$                        & $~\,0.364 \pm 0.083$                 & $~\,10.08 \pm 0.10$                  &                                  \\
 UPM~J0250$-$0559           &                                      & 02:50:40.29                      & $-$05:59:50.89                    & $~\,0.66 \pm 0.11$                                       &                                  & $~\,~\,33.2 \pm 1.0$              & $~\,~\,2.18 \pm 0.85$                        & $~\,0.825 \pm 0.219$                 & $~\,11.08 \pm 0.18$                  &                                  \\
 LP~771--50                 & M5e                                  & 02:56:27.19                      & $-$16:27:38.53                    & $~\,0.29 \pm 0.06$                                       &                                  & $~\,~\,32.6 \pm 0.4$              & $~\,~\,1.61 \pm 0.89$                        & $~\,1.003 \pm 0.373$                 & $~\,11.68 \pm 0.25$                  &                                  \\
 CD$-$44~1173               & K6.5Ve                               & 03:31:55.74                      & $-$43:59:14.92                    & $~\,2.35 \pm 0.20$                                       &                                  & $~\,~\,22.1 \pm 0.0$              & $~\,~\,3.77 \pm 1.09$                        & $-0.875 \pm 0.176$                   & $~\,10.57 \pm 0.13$                  &                                  \\
 UPM~J0409$-$4435           &                                      & 04:09:32.15                      & $-$44:35:38.19                    & $~\,0.59 \pm 0.12$                                       &                                  & $~\,~\,68.1 \pm 0.0$              & $~\,~\,2.34 \pm 0.80$                        & $-0.925 \pm 0.225$                   & $~\,10.59 \pm 0.17$                  &                                  \\
 V833~Tau                   & K2.5Ve                               & 04:36:48.60                      & $+$27:07:55.04                    & $~\,2.58 \pm 0.14$                                       &                                  & $~\,~\,57.1 \pm 0.1$              & $~\,~\,3.82 \pm 1.53$                        & $-0.758 \pm 0.200$                   & $~\,~\,9.67 \pm 0.17$                & G{\"u}d92                        \\
 YZ~CMi                     & M4Ve                                 & 07:44:39.67                      & $+$03:33:01.79                    & $~\,0.72 \pm 0.14$                                       &                                  & $~\,167.0 \pm 0.1$                & $~\,~\,2.36 \pm 1.52$                        & $-0.986 \pm 0.430$                   & $~\,~\,9.64 \pm 0.28$                & Dav78, Vil19                     \\
 G~41--14                   & M3.5V                                & 08:58:56.75                      & $+$08:28:19.76                    &                                                          &                                  & $~\,147.7 \pm 2.0$                & $~\,18.66 \pm 4.03$                          & $-0.895 \pm 0.137$                   &                                      &                                  \\
 MV~Vir                     & K5.5Vkee                             & 14:14:21.16                      & $-$15:21:25.64                    & $~\,3.57 \pm 0.36$                                       &                                  & $~\,~\,34.5 \pm 0.2$              & $~\,~\,1.67 \pm 0.82$                        & $-0.675 \pm 0.294$                   & $~\,~\,9.47 \pm 0.21$                &                                  \\
 G~165--61                  & M4.5Ve                               & 14:17:02.04                      & $+$31:42:44.26                    & $~\,0.71 \pm 0.23$                                       &                                  & $~\,~\,60.0 \pm 2.2$              & $~\,~\,1.96 \pm 0.84$                        & $~\,0.963 \pm 0.278$                 & $~\,10.46 \pm 0.23$                  &                                  \\
 CD$-$38~11343              & M3Ve+M4Ve                            & 16:56:48.49                      & $-$39:05:38.96                    & $~\,1.70 \pm 0.29$                                       &                                  & $~\,~\,63.8 \pm 0.1$              & $~\,~\,7.08 \pm 1.96$                        & $-0.293 \pm 0.096$                   & $~\,10.21 \pm 0.14$                  &                                  \\
 UCAC4~312--101210          &                                      & 17:02:07.98                      & $-$27:40:28.67                    & $~\,0.72 \pm 0.18$                                       &                                  & $~\,~\,41.9 \pm 0.1$              & $~\,~\,3.31 \pm 1.32$                        & $~\,0.744 \pm 0.231$                 & $~\,10.99 \pm 0.20$                  &                                  \\
 Ross~867                   & M4.5V                                & 17:19:52.68                      & $+$26:30:09.94                    & $~\,0.98 \pm 0.16$                                       &                                  & $~\,~\,93.0 \pm 0.1$              & $~\,~\,3.68 \pm 1.12$                        & $~\,0.667 \pm 0.143$                 & $~\,10.07 \pm 0.15$                  & Jac87, Whi89, Qui20              \\
 G~183--10                  & M3.5Ve                               & 17:53:00.26                      & $+$16:54:59.07                    & $~\,0.80 \pm 0.20$                                       &                                  & $~\,~\,44.2 \pm 1.0$              & $~\,~\,2.01 \pm 0.80$                        & $~\,0.531 \pm 0.200$                 & $~\,10.64 \pm 0.20$                  &                                  \\
 SCR~J1928$-$3634           &                                      & 19:28:33.74                      & $-$36:34:39.21                    & $~\,0.43 \pm 0.09$                                       &                                  & $~\,~\,39.5 \pm 0.1$              & $~\,~\,2.44 \pm 1.05$                        & $-0.638 \pm 0.214$                   & $~\,11.35 \pm 0.20$                  &                                  \\
 Ross~776                   & M3.3                                 & 21:16:06.14                      & $+$29:51:52.37                    & $~\,1.12 \pm 0.17$                                       &                                  & $~\,~\,49.2 \pm 0.2$              & $~\,~\,6.26 \pm 1.54$                        & $~\,0.500 \pm 0.094$                 & $~\,10.74 \pm 0.12$                  &                                  \\
 SCR~J2241$-$6119           &                                      & 22:41:44.80                      & $-$61:19:33.24                    & $~\,0.24 \pm 0.02$                                       &                                  & $~\,~\,35.2 \pm 0.0$              & $~\,~\,4.04 \pm 1.09$                        & $~\,0.523 \pm 0.124$                 & $~\,12.18 \pm 0.12$                  &                                  \\
\hline
\multicolumn{10}{c}{Interacting Binaries} \\
\hline
 HR~1099                    & K2Vnk                                & 03:36:47.19                      & $+$00:35:13.52                    & $40.20 \pm 0.90$                                         & Mut84                            & $~\,~\,33.8 \pm 0.1$              & $~\,16.20 \pm 3.51$                          & $-0.161 \pm 0.033$                   & $~\,~\,8.37 \pm 1.61$                & Mut84, Whi95, Sle08, Rav10       \\
 V1154~Tau                  & B6III/IV                             & 05:05:37.70                      & $+$23:03:41.56                    & $~\,34.1 \pm 20.4$                                       &                                  & $~\,~\,~\,3.1 \pm 0.9$            & $~\,~\,2.00 \pm 0.86$                        & $-0.639 \pm 0.201$                   & $~\,~\,9.67 \pm 1.89$                &                                  \\
 $\xi$~UMa                  & F8.5V                                & 11:18:10.18                      & $+$31:31:31.12                    & $29.95 \pm 0.03$                                         & Gri98                            & $~\,114.5 \pm 0.4$                & $~\,~\,1.45 \pm 0.92$                        & $-0.680 \pm 0.302$                   & $~\,~\,6.52 \pm 3.29$                &                                  \\
 BH~CVn~/~HR~5110           & A6m+KIV                              & 13:34:47.90                      & $+$37:10:56.78                    & $17.10 \pm 0.15$                                         & Abb15                            & $~\,~\,21.7 \pm 0.2$              & $~\,~\,6.92 \pm 1.65$                        & $-0.338 \pm 0.078$                   & $~\,~\,9.13 \pm 0.25$                & Mut87, Whi95, Abb15              \\
 V851~Cen                   & K0III                                & 13:44:00.96                      & $-$61:21:58.92                    & $66.33 \pm 0.49$                                         & Kar04                            & $~\,~\,13.5 \pm 0.0$              & $~\,~\,6.85 \pm 2.17$                        & $~\,0.726 \pm 0.154$                 & $~\,~\,8.36 \pm 5.31$                &                                  \\
 KZ~Pav                     & F6V                                  & 20:58:39.76                      & $-$70:25:20.58                    & $17.16 \pm 0.06$                                         & Sur10                            & $~\,~\,~\,8.9 \pm 0.0$            & $~\,~\,1.26 \pm 0.47$                        & $-0.554 \pm 0.201$                   & $~\,~\,9.16 \pm 0.44$                & Sle87, Ste89                     \\
\hline
\multicolumn{10}{c}{Young Stellar Objects} \\
\hline
 $\rho$~Oph~S1              & B3                                   & 16:26:34.11                      & $-$24:23:28.19                    & $~\,12.8 \pm 4.8$                                        & And91                            & $~\,~\,~\,8.2 \pm 0.1$            & $~\,~\,7.92 \pm 1.77$                        & $-0.234 \pm 0.063$                   & $~\,10.89 \pm 0.15$                  & Fal81, And88, And91              \\
 EM*~SR~20                  & G7                                   & 16:28:32.50                      & $-$24:22:45.80                    & $~\,2.39 \pm 0.61$                                       &                                  & $~\,~\,~\,7.4 \pm 0.1$            & $~\,~\,2.12 \pm 0.85$                        & $~\,0.719 \pm 0.215$                 & $~\,11.26 \pm 0.20$                  &                                  \\
\hline
\multicolumn{10}{c}{Magnetic Chemically Peculiar Stars} \\
\hline
 $k^2$~Pup~/~HR~2949                  & B3IV He-W                            & 07:38:49.74                      & $-$26:48:12.84                    & $16.07 \pm 2.92$                                         &                                  & $~\,~\,~\,9.4 \pm 0.8$            & $~\,~\,2.86 \pm 0.81$                        & $~\,0.757 \pm 0.162$                 & $~\,~\,9.52 \pm 0.15$                &                                  \\
 OY~Vel                     & ApSi                                 & 09:01:44.44                      & $-$52:11:19.81                    & $12.44 \pm 4.61$                                         &                                  & $~\,~\,~\,8.8 \pm 0.4$            & $~\,~\,6.87 \pm 1.60$                        & $~\,0.675 \pm 0.116$                 & $~\,10.18 \pm 0.19$                  &                                  \\
 V863~Cen                   & B6IIIe He-S                          & 12:08:05.06                      & $-$50:39:40.79                    & $15.62 \pm 3.00$                                         &                                  & $~\,~\,~\,9.8 \pm 0.3$            & $~\,~\,2.20 \pm 0.63$                        & $~\,0.600 \pm 0.154$                 & $~\,~\,9.39 \pm 0.15$                &                                  \\
 CU~Vir                     & ApSi                                 & 14:12:15.72                      & $+$02:24:34.19                    & $~\,9.62 \pm 3.21$                                       &                                  & $~\,~\,12.6 \pm 0.2$              & $~\,13.72 \pm 3.49$                          & $~\,0.416 \pm 0.074$                 & $~\,10.39 \pm 0.18$                  & Tri00, Let06, Rav10, Lo12        \\
 V1040~Sco                  & B2Ve He-S                            & 15:53:55.82                      & $-$23:58:41.33                    & $~\,9.30 \pm 0.60$                                       & Let18                            & $~\,~\,~\,7.6 \pm 0.4$            & $~\,~\,7.64 \pm 1.75$                        & $~\,0.219 \pm 0.055$                 & $~\,10.61 \pm 0.10$                  & Con98, Mur10, Let18              \\
\hline
\multicolumn{10}{c}{Hot Spectroscopic Binaries} \\
\hline
 HD~32595                   & B8                                   & 05:04:49.06                      & $+$13:18:32.86                    & $10.44 \pm 0.93$                                         &                                  & $~\,~\,~\,3.0 \pm 0.2$            & $~\,~\,3.27 \pm 1.19$                        & $~\,0.618 \pm 0.184$                 & $~\,10.96 \pm 0.16$                  &                                  \\
 Castor~A                   & A1V+dMe                              & 07:34:35.49                      & $+$31:53:14.63                    & $~\,6.57 \pm 0.12$                                       & Tor02                            & $~\,~\,64.1 \pm 3.8$              & $~\,~\,1.30 \pm 0.69$                        & $~\,0.794 \pm 0.295$                 & $~\,~\,8.29 \pm 0.22$                & Sch94, Hel99                     \\
\hline
\end{tabular}
\begin{tablenotes}
  \item[*] $f_p > 0$ corresponds to right handed circular polarisation.
  \item[\textdagger] $R_e$ taken as: $3R_\star$ for single stars as calculated from Gaia DR2 photometry \citep{Gaia2018}, 
    the inter-binary region for interacting binaries, or otherwise from literature where indicated by a reference code.
  \item {\bf Reference codes}:
    Abb15 \citep{Abbuhl2015},
    And88 \citep{Andre1988},
    And91 \citep{Andre1991},
    Con98 \citep{Condon1998},
    Dav78 \citep{Davis1978},
    Fal81 \citep{Falgarone1981},
    Gru12 \citep{Grunhut2012},
    Gri98 \citep{Griffin1998},
    G{\"u}d92 \citep{Gudel1992},
    Hel99 \citep{Helfand1999},
    Jac87 \citep{Jackson1987},
    Kar04 \citep{Karatas2004},
    Let06 \citep{Leto2006},
    Let18 \citep{Leto2018},
    Lo12 \citep{Lo2012},
    Mut84 \citep{Mutel1984},
    Mut87 \citep{Mutel1987},
    Mur10 \citep{Murphy2010},
    Qui20 \citep{Quiroga-Nunez2020},
    Rav10 \citep{Ravi2010},
    Sch94 \citep{Schmitt1994},
    Sle87 \citep{Slee1987},
    Sle08 \citep{Slee2008},
    Ste89 \citep{Stewart1989},
    Tor02 \citep{Torres2002},
    Tri00 \citep{Trigilio2000},
    Sur10 \citep{Surgit2010},
    Vil19 \citep{Villadsen2019},
    Whi89 \citep{White1989b},
    Whi95 \citep{White1995}
\end{tablenotes}
\end{threeparttable}}
  \end{table}
\end{landscape}

\subsection{Cool Dwarfs}

We identify 18 K- and M-dwarf stars within our sample, 15 of which have no previously reported
radio detection. Cool dwarfs typically produce non-thermal radio emission in magnetically
confined coronae. Non-thermal, incoherent gyrosynchrotron emission dominates at
C/X-band, and coherent emission processes at lower frequencies where gyrosynchrotron
emission becomes optically-thick \citep{Gudel2002}. Our cool dwarf detections have fractional
polarisation ranging from $29-100$ per cent. Notes on individual sources are presented below.

\starname{G 131--26} (see Fig.~\ref{fig:cutouts1}) is an M5V flare star demonstrating H$\alpha$
\citep{Hawley1996, Newton2017, Jeffers2018} and ultraviolet \citep{Jones2016} activity, and is
a strong variable X-ray source \citep{Hambaryan1999}.

\starname{CS~Cet} (see Fig.~\ref{fig:cutouts1}) is a spectroscopic binary with a K0IVe primary
which has demonstrated BY Dra type photometric variability due to rotation of starspots
\citep{Watson2000}. This system is a bright ultraviolet and X-ray source \citep{Haakonsen2009,
  Beitia-Antero2016} and is chromospherically active in \ion{Ca}{ii} H and K and H$\alpha$
lines \citep{Isaacson2010}.

\starname{LP~771--50} (see Fig.~\ref{fig:cutouts1}) is an M5Ve-type high-proper-motion star
demonstrating H$\alpha$ line emission \citep{Cruz2002}.

\starname{CD$-$44~1173} (see Fig.~\ref{fig:cutouts1}) is a young K6.5Ve star in the
Tucana-Horologium moving group with rotationally modulated optical variability
\citep{Messina2010}, and has previously exhibited ultraviolet flares \citep{ParkeLoyd2018} and
X-ray activity \citep{Haakonsen2009}.

\starname{V833~Tau} (see Fig.~\ref{fig:cutouts2}) is an extremely active K2.5Ve BY Dra variable
demonstrating both X-ray flares \citep{Fuhrmeister2003} and photometric variability due to
large starspot covering fractions \citep{Olah2001}. \citet{Gudel1992} reported a radio flux of
\SI{5.22+-0.07}{\milli\jansky} at \SI{8.5}{\giga\hertz} for this star which they attribute to
gyrosynchrotron emission due to the observed ${\sim}\SI{e8}{\kelvin}$ brightness
temperature. We detect 76 per cent left circularly-polarised emission with
${S_{888}=\SI{3.94+-0.46}{\mjpb}}$, which corresponds to a brightness temperature of
${\sim}\SI{5e9}{\kelvin}$ and is consistent with optically-thin gyrosynchrotron emission.

\starname{YZ~CMi} (see Fig.~\ref{fig:cutouts2}) is an M4Ve eruptive variable known to produce
bright optical flares \citep{Kowalski2010}, and is a well known radio flare star
\citep{Davis1978, Abada-Simon1997}. Highly circularly-polarised radio bursts have been detected
from this star at centimetre wavelengths \citep{Villadsen2019} attributed to ECM emission from
within extreme coronal plasma cavities which is modulated by stellar rotation. We detect 98 per
cent left circularly-polarised emission, which is consistent with the handedness observed by
\citet{Villadsen2019}.

\starname{G~41--14} (see Fig.~\ref{fig:cutouts2}) is an M3.5V triple star system with recorded
optical flares \citep{Rodriguez-Martinez2020} and H$\alpha$ \citep{Hawley1996, Newton2017,
  Jeffers2018} and ultraviolet \citep{Miles2017} activity, but no previously reported radio
emission. The star was included in a radio interferometric search of nearby stars for planetary
companions \citep{Bower2009} but no detections are reported. A \SI{0.99+-0.14}{\milli\jansky}
FIRST radio source is reported $7\farcs8$ from the RACS position and attributed to an SDSS
galaxy \citep{Thyagarajan2011}. Proper motion correction of G~41--14 to the FIRST epoch results
in a $1\farcs35$ separation from the FIRST position, indicating that the star may have been the
origin of the radio emission. G~41--14 was detected in RASS \citep{Fuhrmeister2003} and is
flagged as displaying X-ray flares.

\starname{MV~Vir~/~HD~124498} (see Fig.~\ref{fig:cutouts2}) is a binary system with a
K5.5Vkee-type primary and an unclassified secondary. The system exhibits strong chromospheric
\ion{Ca}{ii} H and K line emission \citep{Boro2018} and is a bright ultraviolet
\citep{Ansdell2014} and X-ray source \citep{Haakonsen2009}. We detect 68 per cent left
circularly-polarised emission with ${S_{888}=\SI{1.72}{\mjpb}}$ positioned $0\farcs9$ from
HD~124498A and $2\farcs4$ from HD~124489B. Both components are within our positional
uncertainty, so we can not distinguish which component is the source of emission. A
neighbouring, unpolarised radio source is also visible in total intensity, with the combined
source components resembling the morphology of a double-lobed radio galaxy. The high fractional
polarisation of our detection is most easily explained if originating from MV~Vir, and as both
stars are separated $11\farcs6$ from the neighbouring source we believe it is an unrelated
background object.

\starname{G~165--61} (see Fig.~\ref{fig:cutouts3}) is an M4.5V spectroscopic binary that
demonstrates both H$\alpha$ \citep{Newton2017} and ultraviolet \citep{Jones2016}
activity. G~165--61 has a rotational period of 113.6 days \citep{Newton2016}, while M-dwarf
radio bursts are typically associated with young, fast rotators \citep{McLean2012}, and only a
small fraction of $>100$ day period M-dwarfs demonstrate other indicators of magnetic activity
\citep{West2015, Mondrik2018}.

\starname{CD$-$38~11343} (see Fig.~\ref{fig:cutouts3}) is an M3Ve + M4Ve binary system of flare
stars \citep{Tamazian2014}.  The A component has been associated with flaring X-ray emission
\citep{Fuhrmeister2003, Wright2011} and periodic optical variability due to rotation of
starspots \citep{Kiraga2012}. The angular separation between the stars is of the order of our
positional uncertainty at $3\farcs1$, so that it is not clear from which star the emission
originates.

\starname{Ross~867} and Ross~868 (see Fig.~\ref{fig:cutouts3}) are a visual binary system of
M4.5V and M3.5V flare stars with similar stellar properties. Both components have demonstrated
optical flaring \citep{Tamazian2014} and photospheric variability \citep{Kiraga2012},
ultraviolet \citep{Jones2016} and X-ray \citep{Fuhrmeister2003} variability, and H$\alpha$
activity \citep{Hawley1996}. Ross~867 is a well known radio-loud star \citep{Jackson1987,
  White1989b} and in particular has demonstrated radio bursts with moderate circular
polarisation \citep{Quiroga-Nunez2020}, while Ross~868 has never been detected at radio
wavelengths. We detect 67 per cent right circularly-polarised emission with
${S_{888}=\SI{3.80+-0.32}{\mjpb}}$ unambiguously associated with Ross~867, which is separated
from Ross~868 by $16\farcs7$.

\starname{G~183--10} (see Fig.~\ref{fig:cutouts3}) is an M4V-type star with no significant
ultraviolet \citep{Ansdell2014} or H$\alpha$ \citep{Jeffers2018} activity, and no previously
reported radio detection.

\starname{Ross~776} (see Fig.~\ref{fig:cutouts4}) is an M3.3V flare star \citep{Tamazian2014}
showing both H$\alpha$ \citep{Hawley1996, West2015} and ultraviolet activity \citep{Jones2016},
and has previously demonstrated X-ray flaring \citep{Fuhrmeister2003}. This star is a young,
rapid rotator, with a \SI{0.586}{\day} photometric rotation period \citep{Newton2016}.

\vspace*{4pt}\noindent {\bf UPM J0250$-$0559, UPM J0409$-$4435, UCAC4 312$-$101210, SCR
  J1928$-$3634}, and {\bf SCR J2241$-$6119} (see Fig.~\ref{fig:cutouts1}--\ref{fig:cutouts4})
are high proper motion stars without spectral classification, though photometric colour indices
imply these are M-class stars \citep{Lepine2011, Winters2011, Frith2013}. SCR~J2241$-$6119 has
also been proposed as an M7 ultra-cool dwarf from analysis of Gaia colours \citep{Reyle2018}.

\begin{figure*}
  \centering
  \includegraphics[width=6.1in]{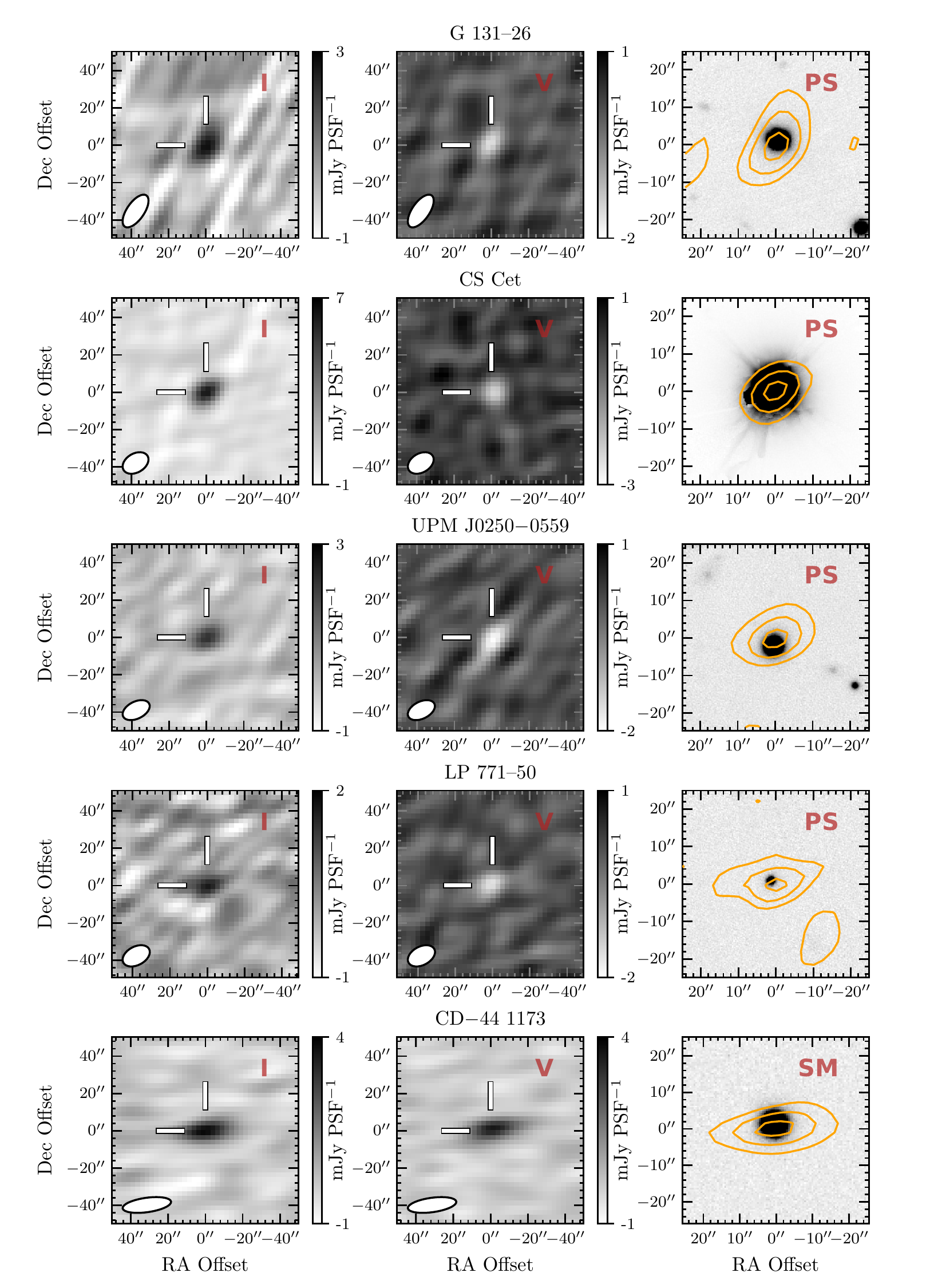}
  \caption{Images of G~131--26, CS~Cet, UPM~J0250$-$0559, LP~771--50, and CD$-$44~1173. RACS
    data is shown in Stokes I (left panels) and Stokes V (middle panels), where positive flux
    density in the Stokes V map corresponds to right handed circular polarisation and negative
    to left handed. The ellipse in the lower left corner of each radio image shows the
    restoring beam. The right panels show Stokes I contours overlaid on optical data from
    Pan-STARRS1 (g-band) or Skymapper for stars above and below $\delta=-30^\circ$
    respectively, with contour levels at 30, 60, and 90 per cent of the peak Stokes I flux
    density. All images have been centred on a frame aligned with the position of the radio
    source, with North up and East to the left. Optical data has been astrometrically corrected
    to the RACS epoch according to the proper motion of the target star. For some of the
    brighter stars the optical data is over-saturated or masked.}\label{fig:cutouts1}
\end{figure*}

\begin{figure*}
  \centering
  \includegraphics[width=6.1in]{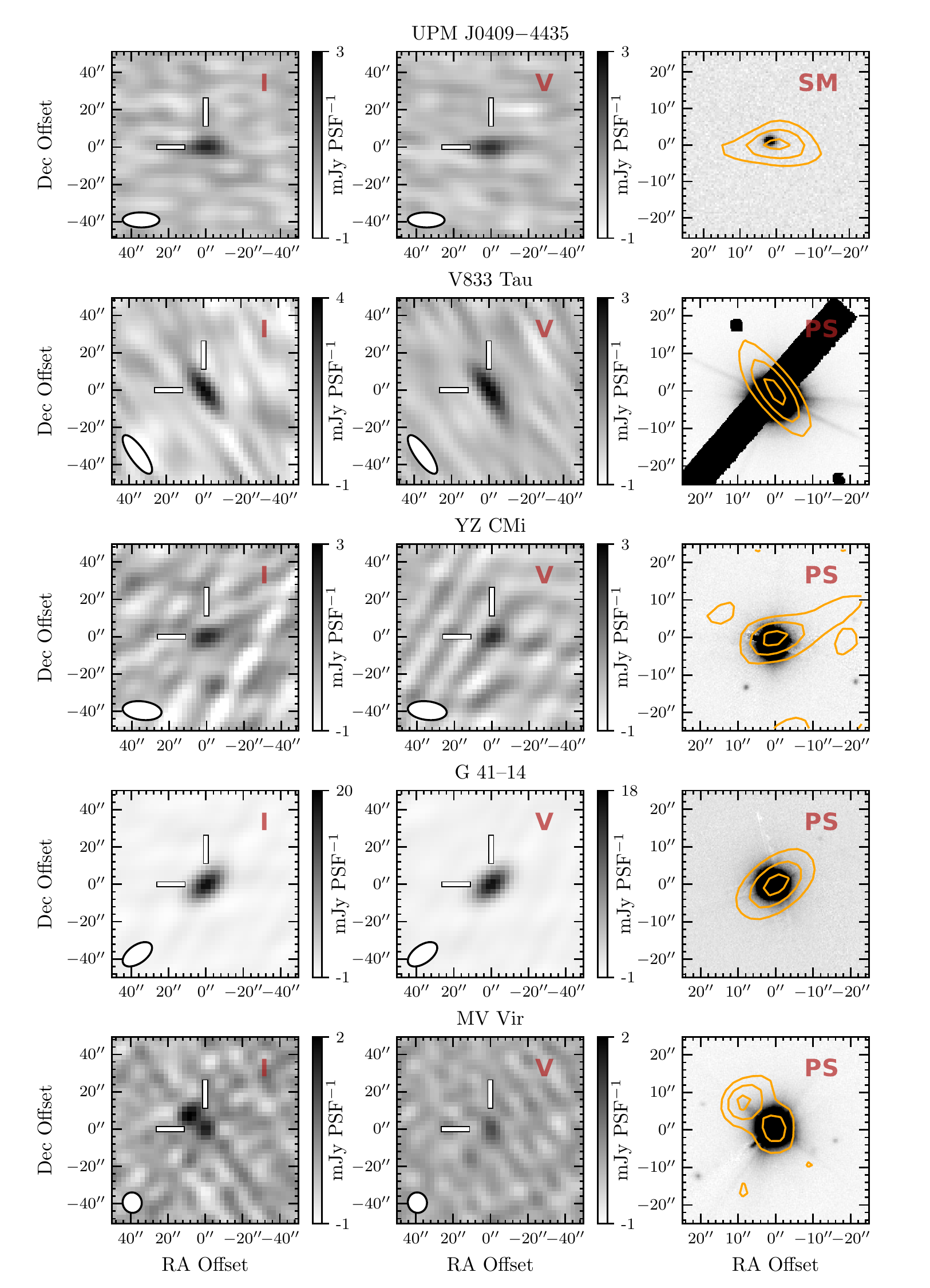}
  \caption{Images of UPM~J0409$-$4435, V833~Tau, YZ~CMi, G~41--14, and MV~Vir. Details as in
    Fig.~\ref{fig:cutouts1}.}\label{fig:cutouts2}
\end{figure*}

\begin{figure*}
  \centering
  \includegraphics[width=6.1in]{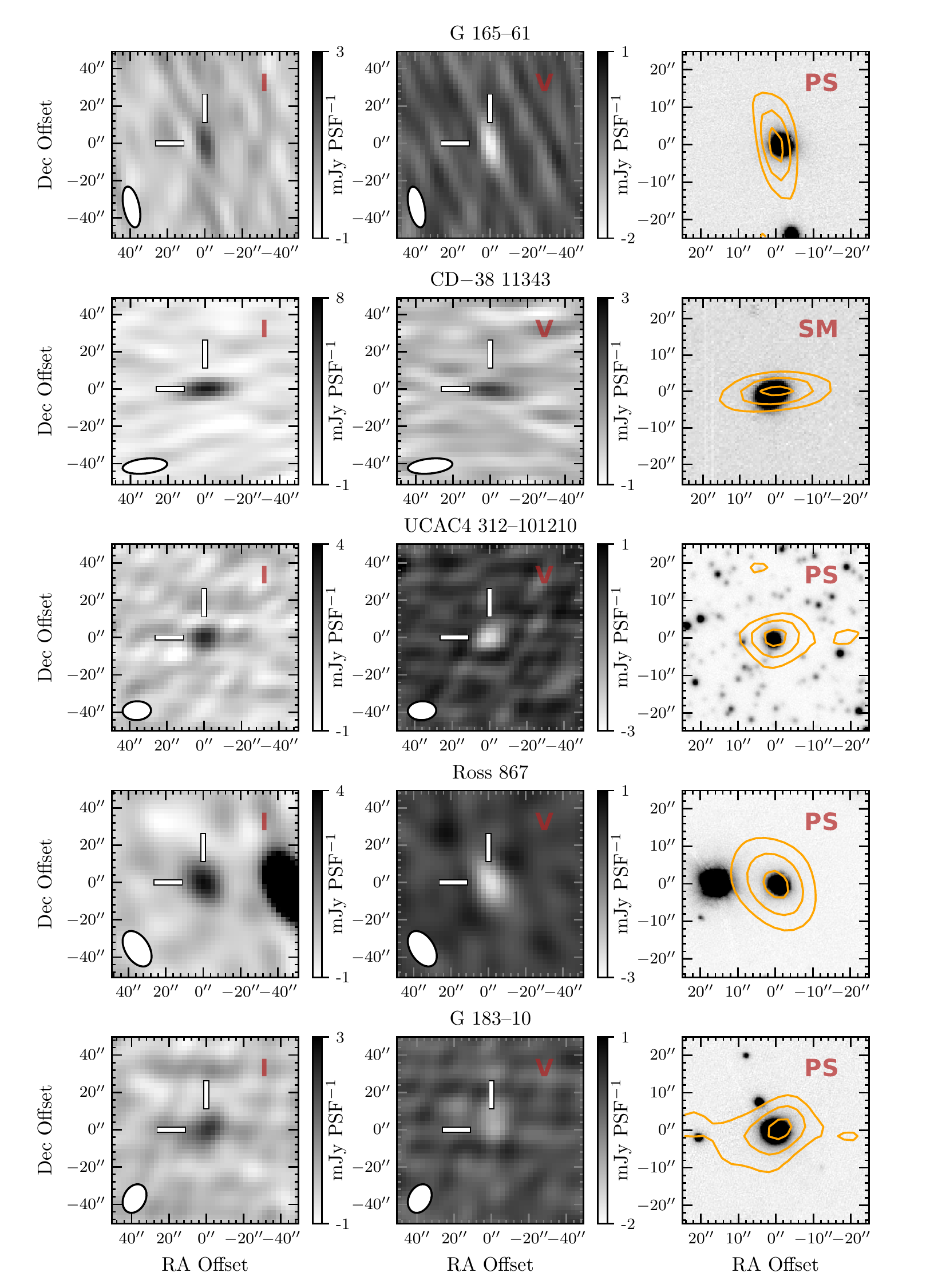}
  \caption{Images of G~165--61, CD$-$38~11343, UCAC4~312--101210, Ross~867, and
    G~183--10. Details as in Fig.~\ref{fig:cutouts1}.}\label{fig:cutouts3}
\end{figure*}

\begin{figure*}
  \centering
  \includegraphics[width=6.1in]{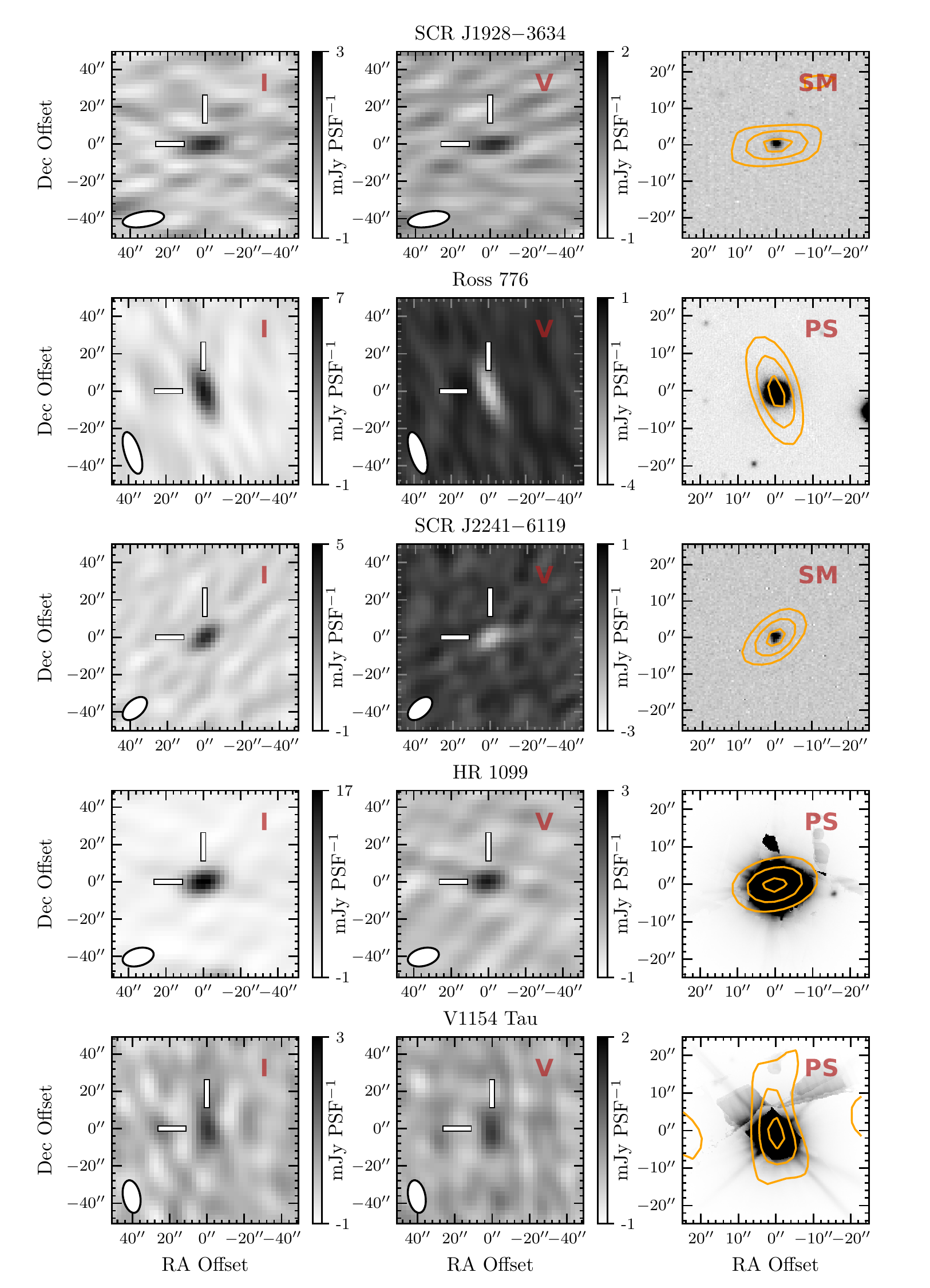}
  \caption{Images of SCR~J1928$-$3634, Ross~776, SCR~J2241$-$6119, HR~1099, and
    V1154~Tau. Details as in Fig.~\ref{fig:cutouts1}.}\label{fig:cutouts4}
\end{figure*}

\begin{figure*}
  \centering
  \includegraphics[width=6.1in]{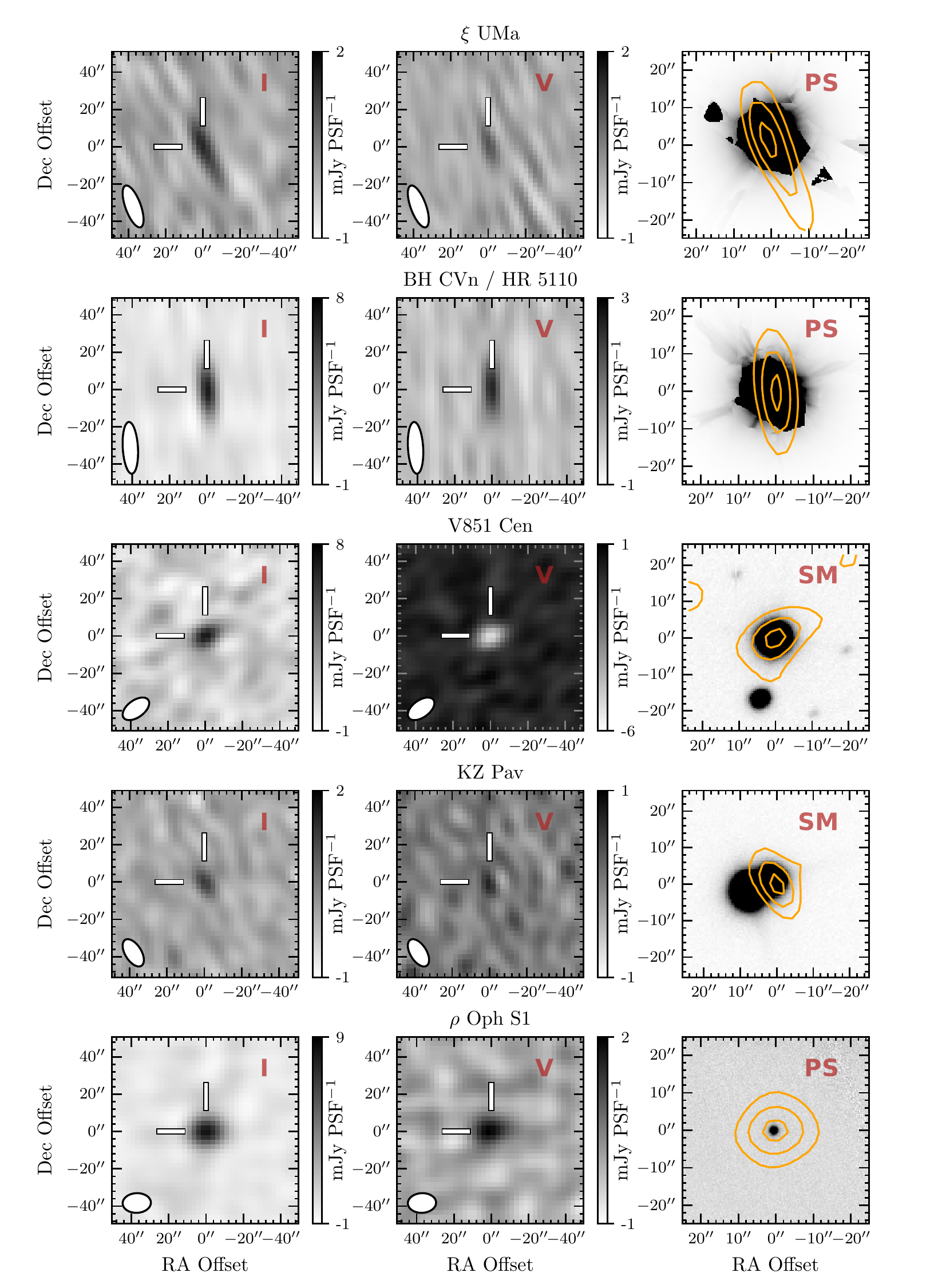}
  \caption{Images of $\xi$ UMa, BH~CVn, V851~Cen, KZ~Pav, and $\rho$~Oph~S1. Details as in
    Fig.~\ref{fig:cutouts1}.}\label{fig:cutouts5}
\end{figure*}

\begin{figure*}
  \centering
  \includegraphics[width=6.1in]{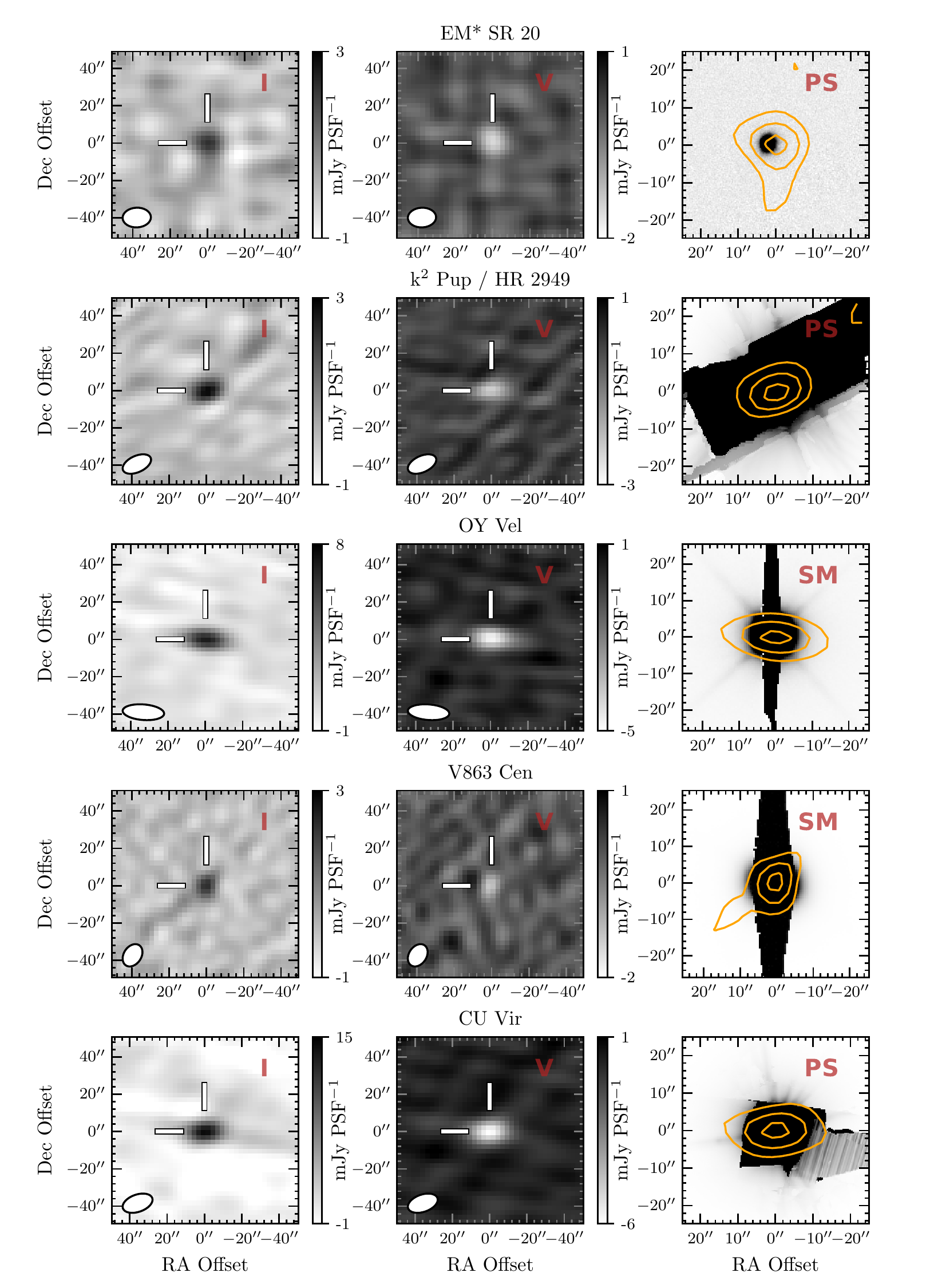}
  \caption{Images of EM*~SR~20, k$^2$~Pup, OY~Vel, V863~Cen, and CU~Vir. Details as in
    Fig.~\ref{fig:cutouts1}.}\label{fig:cutouts6}
\end{figure*}

\begin{figure*}
  \centering
  \includegraphics[width=6.1in]{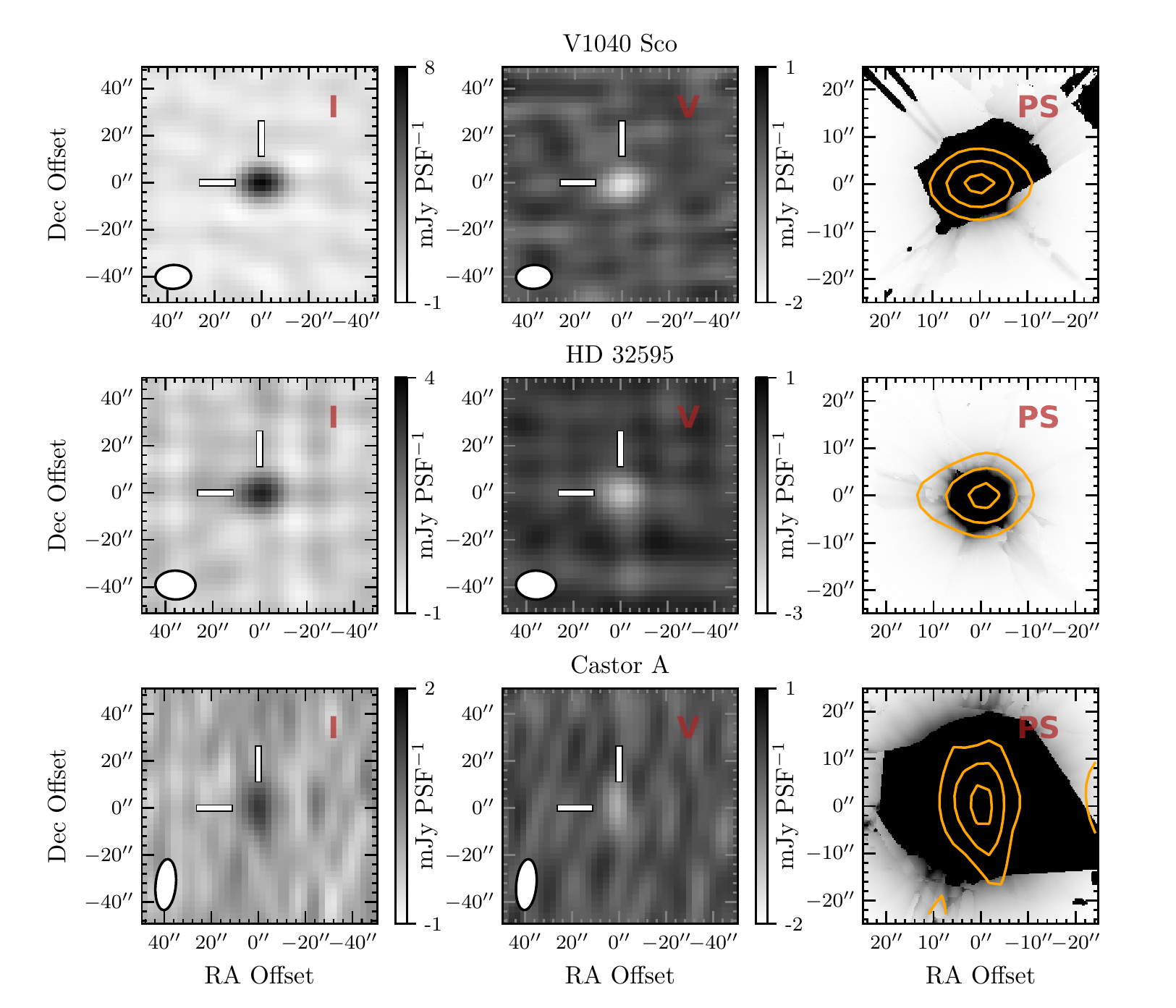}
  \caption{Images of V1040~Sco, HD~32595, and Castor~A. Details as in
    Fig.~\ref{fig:cutouts1}.}\label{fig:cutouts7}
\end{figure*}

\subsection{Interacting Binaries}

We identify 6 interacting binary systems of RS CVn and Algol type, three of which have no
previously reported radio detection. RS CVn and Algol binaries are known to possess strong
magnetic fields generated by rapid, tidally-induced rotation periods. The radio emission
associated with these stars is generally non-thermal with moderate circular polarisation, with
both quiescent gyrosynchrotron emission \citep{Jones1994, Abbuhl2015} and coherent radio bursts
\citep{vandenOord1994, White1995, Slee2008} observed. Our detections in this category have a
fractional polarisation between $16-88$ per cent. Notes on individual sources are presented
below.

\starname{HR~1099~/~V711~Tau} (see Fig.~\ref{fig:cutouts4}) is a K2:Vnk/K4 RS CVn binary with
extensive study at radio frequencies, and is known to exhibit both strong radio flaring and
periods of quiescent emission. VLBI detections have been made during the quiescent period at
\SI{1.65}{\giga\hertz} \citep{Mutel1984} with a source size comparable to the size of the
binary system, and during a flare at \SI{8.4}{\giga\hertz} \citep{Ransom2002} revealing
milliarcsecond structure. HR~1099 is also known to emit highly polarised, coherent radio bursts
which are attributed to ECM driven auroral emission \citep{Slee2008}.

\starname{V1154~Tau} (see Fig.~\ref{fig:cutouts4}) is a B6III/IV-type eclipsing Algol binary
with no previously reported radio detections.

\starname{$\xi$ UMa} (see Fig.~\ref{fig:cutouts5}) is an F8.5:V+G2V RS CVn binary system with
multiple sub-components. The cooler secondary component demonstrates X-ray emission
\citep{Ball2005} attributed to activity induced by a \SI{3.98}{\day} orbit with one of its
sub-components. This system has been frequently targeted by radio observations
\citep{Spangler1977, Morris1988, Drake1989, Drake1992, Bastian2000} but no detections have
previously been reported. The system includes a T8.5 brown dwarf in a wide orbit
\citep{Wright2013}, though this companion is separated by $8\farcm5$ from our radio
detection and cannot explain the observed emission.

\starname{BH~CVn~/~HR 5110} (see Fig.~\ref{fig:cutouts5}) is an F2IV/K2IV binary known to
demonstrate large X-ray flares \citep{Graffagnino1995}, and is one of the most radio active RS
CVn systems. VLBI observations at \SI{15.4}{\giga\hertz} show radio emission originating from
the K-type secondary \citep{Abbuhl2015}, which is typically unpolarised at
\SI{1.5}{\giga\hertz} and consistently right handed at higher frequencies
\citep{White1995}. Non-eclipsing RS CVn systems such as this typically demonstrate a reversal
in the polarisation sense between 1.4 to \SI{5}{\giga\hertz} \citep{Mutel1987}, which is
consistent with our left handed detection at \SI{887.5}{\mega\hertz}. Our detection is 34 per
cent circularly-polarised, slightly above the 20 per cent levels detected at higher frequencies
\citep{White1995}.

\starname{V851 Cen} (see Fig.~\ref{fig:cutouts5}) is a K0III RS CVn binary with chromospheric
activity in both H$\alpha$ and \ion{Ca}{ii} H and K lines \citep{Cincunegui2007}, and has
previously demonstrated X-ray variability \citep{Kashyap1999, Haakonsen2009, Kiraga2012}. The
system has been targeted in two radio surveys due to the chromospheric activity
\citep{Collier1982, Slee1987} but was undetected in both.

\starname{KZ~Pav} (see Fig.~\ref{fig:cutouts5}) is a F6V/K4IV Algol binary of the EA2-type, and
has previously demonstrated variable radio emission at \SI{8.4}{\giga\hertz} with one
\SI{4.8+-1.4}{\milli\jansky} detection in 12 epochs \citep{Slee1987}. The star has also been
observed at \SI{4.9}{\giga\hertz} with persistent \SI{0.3+-0.1}{\milli\jansky} detections over
a 24 hour period \citep{Budding2001}. Both detections correspond to a brightness temperature of
order \SI{e8}{\kelvin} assuming emission originates from the inter-binary region.

\subsection{Young Stellar Objects}

We identify two young stellar objects within our sample. Pre-main sequence stars are
magnetically-active and known to produce both highly polarised coherent emission
\citep{Smith2003} and non-thermal gyrosynchrotron emission \citep{Andre1996}. Non-thermal
emission is less commonly observed in classical T-Tauri (CTT) stars presumably due to
absorption in the ionised circumstellar wind, though exceptions exist associated with extended
magnetospheric structures such as the non-thermal radio knots in the jets of DG~Tau
\citep{Ainsworth2014}. Notes on individual sources are presented below.

\starname{$\rho$~Oph~S1} (see Fig.~\ref{fig:cutouts5}) is a B3 T-Tauri star in the Rho Ophiuchi
cloud complex with a kilogauss globally organised magnetic field, and is a persistent radio
source with a faint $20\arcsec$ wide halo and circularly-polarised core \citep{Falgarone1981,
  Andre1988}. VLBI observation of this star at \SI{4.985}{\giga\hertz} has resolved the core as
an $8-15R_\odot$ region of optically-thin gyrosynchrotron emission with brightness temperature
of order \SI{e8}{\kelvin} \citep{Andre1991}. We detect 23 per cent left circularly-polarised
emission with ${S_{888}=\SI{8.17+-0.28}{\mjpb}}$, which is comparable to the
\SI{5}{\giga\hertz} and \SI{15}{\giga\hertz} measurements by \citealt{Andre1988}.

\starname{EM*~SR~20} (see Fig.~\ref{fig:cutouts6}) is a binary CTT system in the Rho Ophiuchi
cloud complex, with the G7 primary embedded in a protostellar disk \citep{McClure2008}. This
star has been targeted at millimetre wavelengths for cold dust continuum emission, but is
undetected at a \SI{35}{\milli\jansky} limit \citep{Nurnberger1998}.

\subsection{Magnetic Chemically Peculiar Stars}

Compared to later-type stars, magnetically driven, non-thermal emission is less common in hot
stars, presumably due to the lack of an interior convective zone. The exception are the MCP
stars, which have strong, globally organised magnetic fields, and have been observed to produce
both gyrosynchrotron emission driven by equatorial current sheets \citep{Linsky1992}, and
coherent auroral emission \citep{Trigilio2011}. We identify five MCP stars within our sample
with polarisation fractions between $22-70$ per cent. Notes on individual sources are presented
below.

\starname{k$^2$~Pup~/~HR~2949} (see Fig.~\ref{fig:cutouts6}) is a helium-weak B3IV star with
kilogauss surface magnetic fields and helium spectral line variability suggestive of chemically
peculiar spots, in analogue to the cooler Bp/Ap stars \citep{Shultz2015}. There are no reported 
radio detections of this star.

\starname{OY~Vel} (see Fig.~\ref{fig:cutouts6}) is an ApSi magnetic chemically peculiar star
and an $\alpha^2$~CVn variable with no previously reported radio detections.

\starname{V863 Cen} (see Fig.~\ref{fig:cutouts6}) is a chemically peculiar B6IIIe helium-strong
star with kilogauss surface magnetic fields \citep{Briquet2003, Kochukhov2006} and no
previously reported radio emission. Radio detections of Be stars are typically attributed to
interactions between stellar wind outflows and a circumstellar disk resulting in free-free
emission with low fractional polarisation, though a few notable variable detections are
suggestive of a non-thermal mechanism \citep{Dougherty1991, Skinner1993}. Our detection is
inconsistent with free-free emission, with 60 per cent left circularly-polarised emission that is
likely driven by a non-thermal mechanism.

\starname{CU~Vir} (see Fig.~\ref{fig:cutouts6}) is an ApSi $\alpha^2$~CVn variable, and a well
known radio-loud magnetic chemically peculiar star. Quiescent gyrosynchrotron emission with
rotationally modulated variability has been observed from CU Vir \citep{Leto2006}, as well as
two 100 per cent right circularly-polarised pulses that consistently repeat each rotation
period \citep{Trigilio2000, Ravi2010, Lo2012} and are attributed to ECM emission.

\starname{V1040~Sco} (see Fig.~\ref{fig:cutouts7}) is a helium-strong B2.5V star with a
kilogauss global magnetic field, and is the most rapidly rotating magnetic B star known
\citep{Grunhut2012}. This star has been detected at radio frequencies from 1.4 to
\SI{292}{\giga\hertz} \citep{Condon1998, Murphy2010, Leto2018}, with persistent emission
attributed to gyrosynchrotron emission in the rigidly rotating magnetosphere. We measure a
total intensity flux density of \SI{7.87+-0.20}{\mjpb} which is in agreement with that expected
from extrapolation of the gyrosynchrotron spectrum to \SI{887.5}{\mega\hertz}. \citet{Leto2018}
measure circular polarisation of order 10 per cent at \SI{44}{\giga\hertz} decreasing to
order 5 per cent at \SI{6}{\giga\hertz}, undergoing a sign reversal from left handed at high
frequency to right handed below ${\sim}\SI{20}{\giga\hertz}$. This is also in agreement with our
measurement of 22 per cent right handed circular polarisation.

\subsection{Hot Spectroscopic Binaries}

Two stars in our sample are hot, early-type stars that lack identified global magnetic fields
or chemical peculiarities, and which have an unclassified spectroscopic binary
companion. Chemically regular early-type stars are not expected to generate flares due to the
lack of a conventional dynamo \citep{Pedersen2016}, and do not demonstrate other common
features associated with non-thermal, polarised radio emission. Notes on individual sources are
presented below.

\starname{HD~32595} (see Fig.~\ref{fig:cutouts7}) is a spectroscopic binary with a chemically
regular B8 primary, and no previously reported radio detections.

\starname{Castor} (see Fig.~\ref{fig:cutouts7}) is a triple binary system, where Castor~A and B
are both spectroscopic binaries consisting of A-type primaries and dMe secondaries, and
Castor~C is a binary dMe system. The Castor~A system has been detected previously at
\SI{1.4}{\giga\hertz}, \SI{4.9}{\giga\hertz}, and \SI{8.5}{\giga\hertz}
\citep{Schmitt1994}. Our detection is separated by $0\farcs83$ from Castor~A and $3\farcs94$
from Castor~B, so the emission is likely associated with one of the stars in Castor~A.

\section{DISCUSSION}\label{sec:discussion}

\subsection{Emission Mechanism}

\begin{figure*}
  \includegraphics[width=\textwidth]{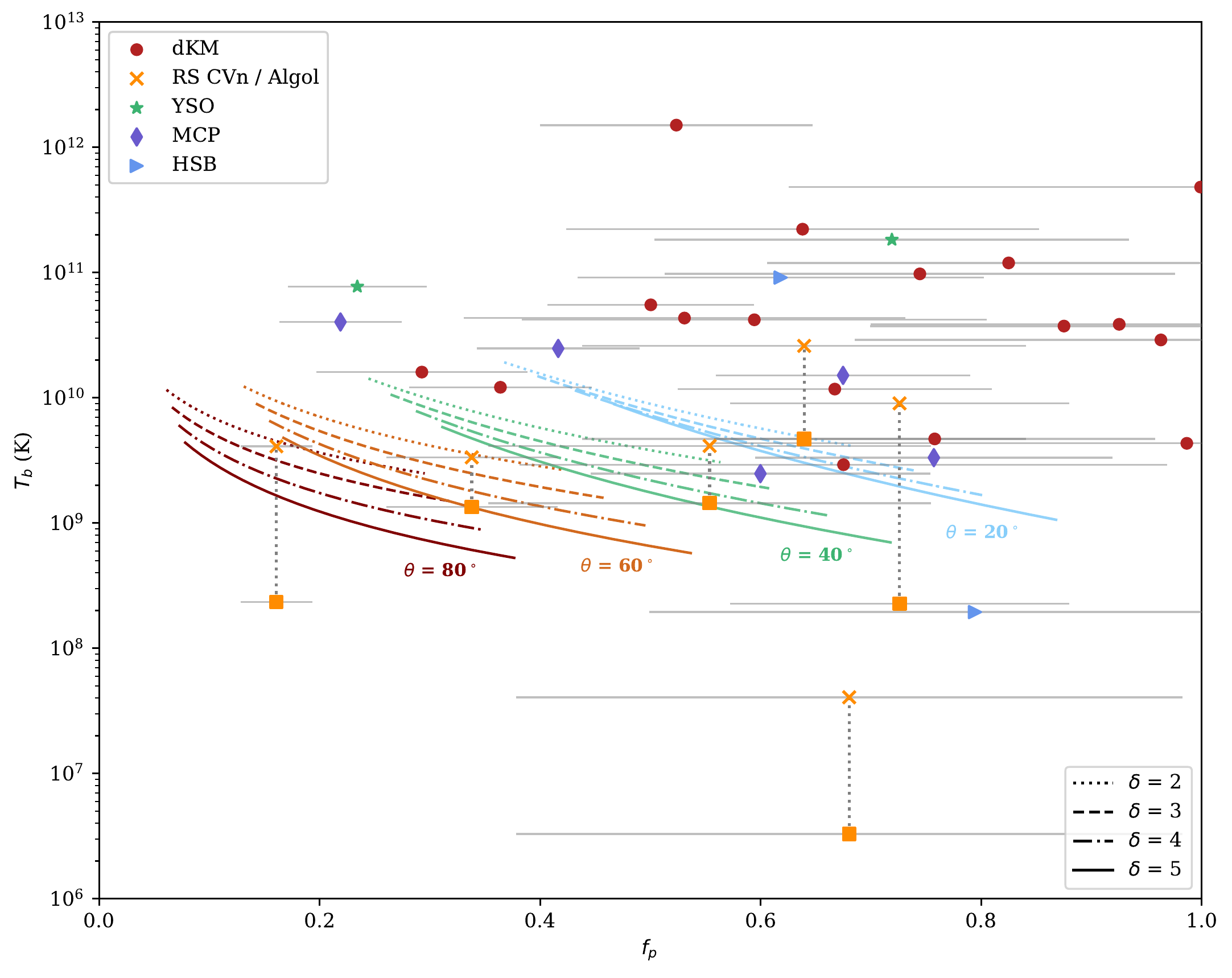}
  \caption{Phase diagram of brightness temperature and fractional polarisation. Brightness
    temperatures correspond to lower limits assuming the observed emission originates from a
    $3R_\star$ disk, with K- and M-dwarfs represented by red circles, YSOs by green stars, MCP
    stars by purple diamonds, and hot spectroscopic binaries as blue right triangles. For RS
    CVn and Algol binaries we show two brightness temperatures: yellow squares represent
    emission originating from the inter-binary region and yellow crosses represent a $3R_\star$
    disk. Empirical models \citep{Dulk1985} are plotted approximating the maximum brightness
    temperature of optically-thin gyrosynchrotron emission for viewing angles of $20^\circ$,
    $40^\circ$, $60^\circ$, and $80^\circ$, and electron power law energy indices of
    $\delta = 2-5$.}\label{fig:tempphase}
\end{figure*}

We calculated the Rayleigh-Jeans brightness temperature of our detections according to
\begin{equation}
  \label{eq:btemp}
  T_B = \frac{S_{888} c^2}{2\pi k_B\nu^2}\frac{D^2}{R_e^2},
\end{equation}
where $k_B$ is the Boltzmann constant, $\nu=\SI{887.5}{\mega\hertz}$ is the observing
frequency, $R_e$ is the length scale of an assumed emission region and $D$ is the stellar
distance.  We have assumed an upper limit to the emission region of $R_e = 3R_\star$ for
detections associated with single stars, and three times the binary separation distance for
interacting binary systems, except where the emission region has been previously determined
with VLBI. As the true emission region may be smaller than our assumed upper limit, the derived
brightness temperature for each of our detections are lower limits, and range from
\SIrange{e6}{e12}{\kelvin}.

In Fig.~\ref{fig:tempphase} we show the position of our detections in a brightness
temperature--fractional polarisation phase space. We show empirical models of the maximum
brightness temperature of optically-thin, non-thermal gyrosynchrotron emission derived from
\citet{Dulk1985} at viewing angles of \SIrange{20}{80}{\degree}, and electron power-law energy
index $\delta$ of $2-5$. Each model is limited to the cyclotron harmonic number range
${10<\omega/\omega_c<100}$, with harmonic number increasing monotonically along the curves from
lower right to upper left.

The fractional polarisation of optically-thick gyrosynchrotron emission is typically less than
20 per cent \citep{Dulk1982} and insufficient to explain the majority of our detections. The RS
CVn and Algol type binary detections are consistent with optically-thin, non-thermal
gyrosynchrotron emission, while the majority of K- and M- dwarf, YSO, and MCP star detections
occupy a region with $T_B$ too high to explain for the observed fractional polarisation, and
are likely driven by a coherent emission process. 

Our upper limit assumptions on the emission region require that the orientation of the magnetic
field varies significantly in the source region, which in turn limits the fractional circular
polarisation due to contributions from both left and right hand polarised emission. Our
detections with high degrees of circular polarisation are therefore likely driven by sources in
a much smaller region with a correspondingly higher brightness temperature, further suggesting
a coherent emission process.

\subsection{Detection Rates}

We detected 33 radio stars within the covered RACS survey area of \SI{34159}{\deg^2} with
detections evenly distributed on the sky, corresponding to a surface density of
${9.66_{-3.01}^{+3.91}\times 10^{-4} \deg^{-2}}$ with errors given by 95 per cent Poisson
confidence intervals. Our survey was limited to $f_p>0.06$ due to the lack of on-axis
polarisation leakage calibration which has now been applied to available RACS data, and we
further restricted our search to positional offsets between Stokes I and V of less than
$2\arcsec$ to reduce contamination from imaging artefacts. Our detection rate is therefore a
lower limit to the surface density of radio stars at \SI{887.5}{\mega\hertz}.

As all detected stars are located within a distance of \SI{150}{pc}, we assume they sample a
Euclidean population of radio stars and follow a cumulative flux density distribution
${N(>S)\propto S^{-3/2}}$. The FIRST survey for radio stars \citep{Helfand1999} detected 26
stars in a survey area of \SI{5000}{\deg^2} above a flux density of \SI{0.7}{\milli\jansky}.
Once scaled to the RACS sensitivity, this corresponds to a surface density of
${2.18_{-1.10}^{+1.73}\times 10^{-3} \deg^{-2}}$ which is comparable to our result.

We extrapolate the number of radio stars detected in this survey to hypothetical future surveys
with ASKAP. Scaling our results to a deep \SI{20}{\micro\jansky\per\beam} RMS survey of the
entire sky south of $\delta=+41^\circ$ implies a surface density of detectable radio stars of
${4.27_{-0.22}^{+0.23}\times 10^{-2} \deg^{-2}}$ and $1400-1500$ total detections. A shallow
but frequent survey covering \SI{10000}{\deg^2} to an RMS of \SI{0.5}{\mjpb} should produce
$6-20$ detections per epoch with a surface density of
${1.17}_{-0.57}^{+0.89}\times 10^{-3} \deg^{-2}$, and would probe the duty cycle and luminosity
distribution of individual variable radio stars.

As the majority of our K- and M-type dwarf detections are inconsistent with incoherent
gyrosynchrotron emission and are likely driven by coherent bursts, we also calculate the
surface density of these detections alone for comparison to previous flare rate estimates in
the literature. We detected 18 K- and M-type dwarfs, assuming the detection of Castor~A can be
attributed to the dMe companion, which results in a surface density of radio-loud cool dwarfs
at \SI{887.5}{\mega\hertz} of ${5.27_{-2.15}^{+3.06}\times 10^{-4} \deg^{-2}}$. In comparison,
\citet{Villadsen2019} report an M-dwarf transient density at \SI{1.4}{\giga\hertz} of
\SI{2.26e-3}{\deg^{-2}} once scaled to the RACS sensitivity, which implies RACS should produce
${\sim}80$ M-dwarf detections. This difference may reflect an increase in cool star radio
activity at \SI{1.4}{\giga\hertz}, or the selection of six highly active M-dwarfs in the
\citet{Villadsen2019} study which are not representative of radio flare rates in the larger
population.

\section{CONCLUSIONS}

We have completed the first all-sky circular polarisation search for radio stars at centimetre
wavelengths within the Rapid ASKAP Continuum Survey, identifying 10 known radio stars and a
further 23 with no previous radio detection. Our sample includes late-type dwarfs, interacting
and chromospherically active binaries, young stellar objects, and magnetic chemically peculiar
stars, demonstrating the variety of magnetically-active stars detectable with this
technique. Many of our detections are highly polarised with brightness temperatures that are
inconsistent with an incoherent emission mechanism. These stars are attractive targets for
followup observations to determine the emission mechanism and further constrain the
magnetospheric properties of the emission environment.

This survey represents a sample of polarised radio stars without any other selection bias and
implies a lower limit to the surface density of radio stars above \SI{1.25}{\milli\jansky} at
\SI{887.5}{\mega\hertz} of ${9.66_{-3.01}^{+3.91}\times 10^{-4} \deg^{-2}}$. Application of
on-axis polarisation calibration to publicly released ASKAP data will allow lower fractional
polarisation and fainter emission to be detected, and extension of this search technique to future
ASKAP surveys will produce a significantly larger sample and the potential for detections in
multiple epochs. These observations will present an opportunity to determine improved
population statistics for the stellar parameters associated with non-thermal radio emission,
and to study burst rates and energetics for thousands of individual stars.

\section*{Acknowledgements}
We thank the anonymous referee for feedback that strengthened this work. We also thank Phil
Edwards and Alec Thomson for helpful comments and suggestions. TM acknowledges the support of
the Australian Research Council through grant FT150100099. JP, AZ, and JKL are supported by
Australian Government Research Training Program Scholarships. DLK was supported by NSF grant
AST-1816492. The Australian Square Kilometre Array Pathfinder is part of the Australia
Telescope National Facility which is managed by CSIRO. Operation of ASKAP is funded by the
Australian Government with support from the National Collaborative Research Infrastructure
Strategy. ASKAP uses the resources of the Pawsey Supercomputing Centre.  Establishment of
ASKAP, the Murchison Radio-astronomy Observatory and the Pawsey Supercomputing Centre are
initiatives of the Australian Government, with support from the Government of Western Australia
and the Science and Industry Endowment Fund. We acknowledge the Wajarri Yamatji as the
traditional owners of the Murchison Radio-astronomy Observatory site. The International Centre
for Radio Astronomy Research (ICRAR) is a Joint Venture of Curtin University and The University
of Western Australia, funded by the Western Australian State government. Parts of this research
were supported by the ARC Centre of Excellence for All Sky Astrophysics in 3 Dimensions (ASTRO
3D), through project number CE170100013. This research made use of the following {\sc python}
packages: {\sc Astropy} \citep{Astropy2013, Astropy2018}, a community-developed core Python
package for Astronomy, {\sc matplotlib} \citep{Hunter2007}, a Python library for publication
quality graphics, {\sc NumPy} \citep{vanderWalt2011, Harris2020}, and {\sc pandas}
\citep{McKinney2010, McKinney2011}.

\section*{Data Availability}
The data analysed in this paper are accessible through the CSIRO ASKAP Science Data 
Archive \citep[CASDA;][]{Chapman2017} under project code AS110. Note that the images used in 
this paper may have properties that differ slightly from the publicly released versions.

\bibliographystyle{mnras}
\bibliography{bibfile}

\bsp	
\label{lastpage}
\end{document}